\documentclass[aps,pra,amsfonts,floatfix,superscriptaddress, twocolumn, 10pt]{revtex4-2}
\usepackage[utf8]{inputenc}
\usepackage{amssymb,amsmath}
\usepackage{bm}
\usepackage{multirow, array}
\usepackage{hyperref, float}
\usepackage{graphicx}
\usepackage[normalem]{ulem}
\usepackage{xcolor}

\newcommand{\refFig}[2][]{Fig.~\ref{#2}#1}
\newcommand{\refEq}[1]{Eq.~(\ref{#1})}

\newcommand{\refSec}[1]{Sec.~\ref{#1}}

\newcommand{\refTab}[1]{Tab.~(\ref{#1})}

\renewcommand{\vec}[1]{\mathbf{#1}}

\begin{document}

\title{Finite-size effects in Schulz-Shastry-Luttinger 
models\\ for determining anyonic signatures in 1d spin chains}
\author{B. Perković}
\affiliation{Physics Department,  Massachusetts Institute of Technology,  182 Memorial Dr, Cambridge, MA 02139, USA}
\author{M. Bonkhoff}
\affiliation{I. Institute for Theoretical Physics, Universit{\"a}t Hamburg, Notkestraße 9, 22607 Hamburg, Germany}
\author{T. Posske}
\affiliation{I. Institute for Theoretical Physics, Universit{\"a}t Hamburg, Notkestraße 9, 22607 Hamburg, Germany}

\begin{abstract}
We study finite-size properties of Schulz-Shastry-Luttinger liquids to reveal anyonic signatures, realized as low-energy excitations on top of the helical ground state in saturated spin-1/2 zigzag chains. The model features asymmetric and marginal couplings of density and phase gradients and belongs to the Schulz-Shastry class. We investigate periodic and Dirichlet boundary conditions and discuss its diagonalization as well as its stability. Although Dirichlet boundary conditions require a fine-tuning of coupling constants and universal parameters, only their magnitude is restricted for cyclic systems. We derive boundary characteristic quantities like Friedel oscillations and persistent currents. Finally, we discuss the bulk and boundary behavior of the longitudinal spin correlations including subleading corrections.

    \end{abstract}

\maketitle

In one spatial dimension (1d), Luttinger liquid behavior is omnipresent and largely determines the properties of quantum liquids at low energies, or of critical phases \cite{Haldane_1981,Haldane1981,HALDANE1981153,sénéchal1999introductionbosonization,Giamarchi2003,eggert2009,Cazalilla2011}. The universal applicability of the approach for canonical particles, or spin degrees of freedom, is rooted in a characteristic property of one-dimensional systems, Bose-Fermi duality, and the concept of statistical transmutation. Beyond the exact relation of Tonks and Girardeau between free fermions and bosons with infinitely strong on-site interactions \cite{Tonks1936,Girardeau1960}, or the operator identities of Jordan and Wigner \cite{Jordan1928}, this describes in a wider sense the dependence of exchange statistics on interactions in $1d$ and the possibility to mediate between different statistics via string or disorder operators \cite{Posske2017,Fradkin2017,Valentirojas2020,Valentirojas2023,ValentíRojas2025}. As such, a variety of lattice models with inherently different exchange statistics can have the same universal low-energy fixed point, whereas their differences reside exclusively in the microscopic origins of universal parameters, statistically non-invariant observables, and in the zero modes for finite systems \cite{Haldane_1981,Haldane1981,HALDANE1981153,Cazalilla_2004,Cazalilla2011}.
\\
\\
Beyond canonical particles and spin degrees of freedom, abelian Tomonaga-Luttinger liquids can also host anyonic statistic and fractional excitations \cite{Wu1995,Pham_2000,Calabrese2007,Bellazzini_2011,LIGUORI2000577}. Realizations of such low-energy theories are found in the continuum, i.e., for extensions of Kundu anyons \cite{Kundu1999,Batchelor_2008,Frolian2022,Tarurell2022}, and in the lattice for anyonic Hubbard models, which have recently been experimentally realized \cite{Greschner2014,Eckardt2016,BonkhoffPosske2021,Sen2021,Kwan2024,Naegerl2025,Kwan2026}. For this type of models, exchange statistic of the low-energy degrees of freedom appears intra-channel, i.e., within each degree of freedom. 
Yet, there exists another class of models that hosts fractional excitations in the Luttinger regime which were originally studied for fermionic systems subject to inter-species gauge potentials or correlated hopping processes \cite{Schulzshastry1998,OSTERLOH2000531}. For the second type of models, fractional exchange statistic appears between degrees of freedom rather than within. Alternatively, these inter-channel anyons can be found in highly saturated spin zigzag chains \cite{Batista2009,Batista2012,Batista2014}, or related bosonic models at strong coupling \cite{Santos2014}. In the absence of external fields, these special spin chains host an exactly solvable ground space spanned by helical spin excitations as well as a condensate of $q$-deformed particles, dubbed hardcore anyon \cite{Batista2009,Batista2012,Batista2014,Posske2019,Gerken2023,Gerken2025,Shiraishi2025,Gerken2026}.
External fields polarize the spin-chain, inducing dilute magnonic excitations around the two helical wavevectors $\pm Q$. The interactions between these magnons are purely statistical, thus the model can be recast in a free anyonic liquid by a Jordan-Wigner transformation \cite{Schulzshastry1998,Batista2014}.
The connection of particle statistics in Luttinger liquids with their zero mode structure warrants theoretical understanding of finite-size effects for such models.
\\
\\
Here, we provide exact solutions for such Schulz-Shastry-type models with open and periodic boundary conditions. Although the statistical coupling constants for the cyclic case are only restricted in magnitude, the open case requires additional fine tuning to ensure conformal invariance at the boundary. This fine-tuning ensures that the flux-tube attachment that induces statistical transmutation is a canonical transformation, where the coupling constants inherit the role of a statistical phase in analogy to the seminal work of Leinaas and Myrheim on 1d anyons. For the experimentally relevant case of the spin 1/2 zigzag chain, we reveal statistical transmutation by means of the simplest, statistically invariant observable in the context of one-dimensional anyons, Friedel oscillations of the local, longitudinal spin operator. There, we additionally find different modulations of these oscillations for even or odd numbers of excitations above the ground state. For the periodic system, on the other hand, the finite-size spectrum is intertwined, coupling density and current zero mode inextricably. Here, we find unconventional persistent current patterns for a statistical parameter of rational multiples of $\pi$, which are created by the interplay of parity with commensurability effects in the total number of excitations. Finally, we exemplarily discuss the bulk-boundary behavior of operators at the hand of the longitudinal correlation function of the spin chain including subleading corrections. We find a chiral imbalance in the subleading corrections, while the full correlation function is invariant with respect to parity and complex conjugation. This chiral imbalance can be probed by spectroscopic measures.
\\
\\
We introduce the general model and concepts in \refSec{sec:introduction}, and diagonalize the Schulz-Shastry Hamiltonian for finite system lengths in \refSec{sec: diagonalization}, using zero-mode bosonization techniques. We investigate Dirichlet and cyclic boundaries without the loss of generality, but restrict ourselves to the realization in the zigzag chain \cite{Batista2009} to discuss the physical implications.   
In \refSec{subsec: persistentcurrents} we analytically investigate persistent currents for statistical parameters that are rational multiples of $\pi$. For Dirichlet boundaries, on the other hand, we discuss the impact of statistical transmutations on Friedel oscillations in \refSec{subsec: friedeloscillations}. Finally in \refSec{subsec: friedeloscillations}, we exemplarily discuss longitudinal spin correlations, and their bulk and boundary behaviour, to reveal the chiral imbalance of subleading contributions as a characteristic anyonic signature.

\section{Model}
\label{sec:introduction}
We consider a two-chain Luttinger liquid of Schulz-Shastry type, consisting of asymmetric, inter-chain coupling of dual fields \cite{Schulzshastry1998,Pham_2000,Batista2014}. The Hamiltonian density $\hat{h}(x)$ consists of two independent Tomonaga-Luttinger liquids coupled by off-diagonal interactions that are parametrized by $g_{1,2}$\cite{sénéchal1999introductionbosonization,Giamarchi2003,Cazalilla_2004,eggert2009,Cazalilla2011}, 
\begin{align}
     \hat{H}=&\int_{-\infty}^{\infty} dx\phantom{a} \hat{h}(x)\label{Hamiltonian General PBC}
    \\
    \hat{h}(x)=&\sum_{j=1,2}\frac{v_j}{2\pi}\left[\frac{1}{K_j}\left(\partial_{x} \hat{\phi}_j(x)\right)^{2}+K_j \left(\pi \hat{\Pi}_j(x)\right)^{2}\right]\nonumber\\
 +& \ \frac{g_1}{4\pi} \partial_{x} \hat{\phi}_1(x) \hat{\Pi}_2(x)+ \frac{g_2}{4\pi} \partial_{x} \hat{\phi}_2(x) \hat{\Pi}_1(x)\nonumber.
\end{align}
The fields $\hat{\phi}_j(x)$ and $\hat{\Pi}_j(x)=\frac{1}{\pi}\partial_x\hat{\theta}_j(x)$ are canonically conjugated to each other
\begin{align}
    \left[\hat{\phi}_j(x),\hat{\Pi}_k(y)\right]= i \delta_{jk}\delta(x-y),
\end{align}
and are compactified, i.e., they correspond to arguments of vertex operators that relate physical entities to low-energy degrees of freedom \cite{sénéchal1999introductionbosonization,Cazalilla_2004}.
Both chains $j=1,2$ are parametrized by their universal constants, i.e., the  Luttinger parameters $K_j$ and the sound velocities $v_j$, while the inter-chain couplings $g_1$ and $g_2$ are marginal and break the individual reflection symmetries of the dual fields \cite{Schulzshastry1998,Pham_2000,OSTERLOH2000531}. 
Provided that $\hat{H}$ is bounded from below, the bosonic Hamiltonian is stable and can be diagonalized by a Fourier decomposition of the dual fields and subsequently solving the linear system of equations for the coefficients \cite{Giamarchi2003}. The solutions are eigenmodes with linear dispersion and velocities $\tilde{v}_{1,2} $,
\begin{align}
& \tilde{v}_{1,2} = \sqrt{\frac{1}{2}\left(T\pm\sqrt{T^2-4\delta}\right)},\label{eq: soundvelocitiesbulk}
\\
&T=\bar{v}^2_1+\bar{v}^2_2+(v_1q_1+v_2q_2)^2\label{trace0bulk},
\\
&\delta=\bar{v}^2_1\bar{v}^2_2,\label{determinant0bulk}
\end{align}
where
\begin{align}
&\bar{v}_j=v_j\sqrt{1-q^2_{j}},\quad  q_j=\frac{g_j}{4\pi\sqrt{v_jv_{i \neq j}}}\sqrt{\frac{K_j}{K_{i \neq j}}},
\end{align}
and $j=1,2$. From the non-negativity of the Hamiltonian density $\hat{h}(x)$, i.e., a quadratic form in density and phase gradients, we find that the coupling constants $g_{1,2}$ are restricted in magnitude,
\begin{align}
 \vert q_j\vert\leq1 \label{stabilityfluxtube},
\end{align}
Furthermore, \refEq{trace0bulk} shows that for the specific fine-tuning of parameters, i.e., 
\begin{align}
    q_1=-\frac{v_2}{v_1}q_2,\label{Dirichletboundaryissue3}
\end{align}
the original degrees of freedom simply decouple and propagate independently with renormalized velocities $\bar{v}_j$. At the operator level, this fine-tuning amounts to the fact that the so-called flux-tube attachment becomes a canonical transformation \cite{Jordan1928,Batista2014,Pham_2000,Valentirojas2020,Valentirojas2023,ValentíRojas2025},
\begin{align}
    &\tilde{\phi}_j(x)\equiv\hat{\phi}_j(x),\quad\alpha_j =\pi q_{i\neq j} \sqrt{\frac{v_{i\neq j} }{ v_j K_{i\neq j}K_j}}\label{fluxtubeattachementphi}
\\
    &\tilde{\Pi}_{j}(x)=\frac{1}{\pi}\partial_x\tilde{\theta}_j(x)\overbrace{=}^{\refEq{Dirichletboundaryissue3}}\hat{\Pi}_{j}(x)+\frac{\alpha_{j}}{\pi^2}\partial_x\hat{\phi}_{i\neq j}(x).
\label{fluxtubeattachement}
\end{align}
Consequently, the density stiffness of one channel, i.e., $\frac{\bar{v}_j}{\bar{K}_j}$ is decreased via the inter-channel coupling,
\begin{align}
 \hat{H}=&\int_{-\infty}^{\infty} dx\phantom{a} \sum_{j=1,2}\hat{h}_j(x),\label{decoupledLL0}
\\
h_j(x)=&\frac{v_j}{2\pi}\left[\frac{(1-q^2_{ j})}{K_j} \left(\partial_{x} \tilde{\phi}_j(x)\right)^{2}+K_j\left(\pi \tilde{\Pi}_j(x)\right)^{2}\right].\nonumber
\end{align}
When the inequality in \refEq{stabilityfluxtube} is saturated, i.e., $q_j=1$, the free boson limit of Luttinger liquids is obtained if $v_j$ and $K_j$ remain finite \cite{Giamarchi2003,Cazalilla_2004}. 
\\
\\
Positive semi-definiteness is not the only requirement on the Hamiltonian, and in the presence of a boundary  \refEq{Dirichletboundaryissue3} becomes essential. In order to show this, it is instructive to investigate \refEq{Hamiltonian General PBC} on a semi-infinite line, i.e.,
\begin{align}
     \hat{H}=&\int_{0}^{\infty} dx\phantom{a} \hat{h}(x).\label{eq: hamiltoniandensity}
\end{align}
In analogy to the classical wave equation \cite{Alembert1749}, we require the density field $\hat{\phi}_j(x)$ to vanish at the boundary $x=0$ for all times , 
\begin{align}
    \hat{\phi}_j(0)=0, 
\end{align}
which consequently must be true for its derivatives. The first derivative, and subsequently all higher orders, vanish exactly if the the flux-tube attached phase variable $\tilde{\theta}_j(x)$ obeys Neumann boundary conditions,
\begin{align}
    &\partial_t\hat{\phi}_j(0)=v_jK_j\pi\hat{\Pi}_j(0)+\frac{g_{i\neq j}}{4\pi}\partial_x\hat{\phi}_{i\neq j}(0)=0.\label{firstderivativeboundary}
\end{align} 
Likewise, the second order Heisenberg equation only reduces to the ordinary wave equation if \refEq{Dirichletboundaryissue3} is met,
\begin{align}
    &\partial^2_t\hat{\phi}_j(0)=\tilde{v}^2_j\partial^2_x\hat{\phi}_j(0)\nonumber
    \\
    &\quad\quad\quad\quad+\frac{v_1K_1g_1+v_2K_2g_2}{4\pi}\partial^2_x\tilde{\theta}_{i\neq j}(0)=0.\label{secondderivativeboundary}
\end{align}
In conclusion, the fine-tuning condition in \refEq{Dirichletboundaryissue3} ensures conformally invariant boundary conditions on the semi-infinite line, analogously to the periodic extension of the d'Alembert solution from $\mathbb{R}^+$ to $\mathbb{R}$ \cite{Alembert1749,CARDY1989581,cardy1991a,Recknagel_Schomerus_2013}.
\\
\\
Originally, such finely tuned models were considered for fermions subject to gauge potentials \cite{Schulzshastry1998}. However, alternative physical realizations exist in a nearly-saturated spin $1/2$ zigzag chain \cite{Batista2009,Batista2012,Batista2014} or in related, strongly-coupled bosonic models \cite{Santos2014}, i.e.,
\begin{align}
 &K_1=K_2=1,  \label{kbatista}
 \\
 &v_1=v_2=v,\label{vbatista}
 \\
 &g_1=g=-g_2,\label{gbatista}
\end{align} 
with statistical parameter
\begin{align}
 &\alpha=\frac{g}{4 v} . \label{alphabatista}
\end{align}
The fields describe symmetric $\hat{\phi}_j$ and anti-symmetric $\hat{\theta}_j$ linear combinations of chiral excitations around a reference state \cite{Giamarchi2003}. However, the Fermi points here correspond to the pitch angles $\pm Q$ of two counter-rotating spin helices \cite{Batista2009,Batista2012,Batista2014}, i.e., 
\begin{align}
&\hat{S}^-_j =\hat{c}_j e^{i\pi\sum_{l<j}\hat{n}_j},\label{batistaderivation}\phantom{a}\hat{\Psi}^{R,L}_{j}(x)\propto e^{i(\pm\hat{\phi}_j(x)-\hat{\theta}_j(x))},
\\
     &\hat{c}_j =    \sum_{j=1,2,\eta=R,L}  e^{i(-1)^{j+1}Qx}\hat{\Psi}^{\eta}_{j}(x), \label{lowenergyürojection}
     \\
     &Q=\arccos\left(-\frac{J_1}{4J_2}\right)\label{helicalangle}.
\end{align}
The parameters in \refEq{kbatista}--\refEq{gbatista} obey \refEq{Dirichletboundaryissue3}
and the flux-tube attachment in \refEq{fluxtubeattachement}  corresponds to a Jordan-Wigner string at the level of vertex operators $\hat{\Psi}^{R,L}_{j}(x)$ \cite{Jordan1928,Batista2014,Pham_2000,Valentirojas2020,Valentirojas2023,ValentíRojas2025}. By this, the transformed constituents of \refEq{lowenergyürojection} obey inter-mode anyonic exchange statistics with statistical parameter $\alpha$ according to \refEq{alphabatista}, while the anti-commutation relations of the physical fermions $ \hat{c}_j$ remain intact \cite{Pham_2000},
\begin{align}
    &\tilde{\Psi}^{\eta}_j(x)=\hat{\Psi}^{\eta}_j(x)e^{i\frac{\alpha}{\pi}\tilde{\phi}_{\tilde{j}\neq j }(x)},\label{jordanwigner}
    \\
    &\tilde{\Psi}^{\eta}_j(x)\tilde{\Psi}^{\tilde{\eta}}_{\tilde{j}}(\tilde{x})+e^{-i\alpha\operatorname{sgn}(x-\tilde{x})}\tilde{\Psi}^{\tilde{\eta}}_{\tilde{j}}(\tilde{x})\tilde{\Psi}^{\eta}_j(x)=0,\phantom{a}j\neq \tilde{j}
    \\
    &\tilde{\Psi}^{\eta}_j(x)\tilde{\Psi}^{\tilde{\eta}}_{j}(\tilde{x})+\tilde{\Psi}^{\tilde{\eta}}_{j}(\tilde{x})\tilde{\Psi}^{\eta}_j(x)=0.
\end{align}
The transformation in \refEq{jordanwigner} complicates the calculation of non-diagonal correlation functions as the cost of a simpler, decoupled Hamiltonian \cite{Pham_2000,Batista2014}. Reconsidering \refEq{firstderivativeboundary} from the perspective of statistical transmutation, allows us to draw the analogy between $\alpha$ and the statistical parameter of Leinaas and Myrheim that ensures a zero probability current at coincident points in the configuration space \cite{Leinaas1977,Posske2017}. Here, the probability current corresponds to the temporal derivative of the fields $\hat{\phi}_j$ at the boundary according to \refEq{firstderivativeboundary}.

\section{Diagonalization}
\label{sec: diagonalization}
The considerations in \refSec{sec:introduction} allow us to draw conclusions about the stability of the bulk Hamiltonian and its conformal invariance in the presence of a single boundary. In order to investigate an actual finite system, on the other hand, we need to consider the ground state of the system, not only its excitations  described in \refSec{sec:introduction}. The ground state energy scales with the inverse system length $L^{-1}$ and is described by zero modes that determine the conformal spectrum. In the following section, we discuss the diagonalization of periodic boundary conditions as well as Dirichlet boundary conditions by means of zero mode bosonization techniques \cite{Haldane_1981,sénéchal1999introductionbosonization,Giamarchi2003,Cazalilla_2004,Cazalilla2011}.

\subsection{Periodic Boundary Conditions }
\label{subsec: Diagonalizationperiodic}
For periodic boundary conditions, we require that the fields $\hat{\phi}_j(x)$ and $\hat{\theta}_j(x)$ are
identified at the positions $x$ and $x+L$. As these fields are compactified on a circle, they are defined up to the shifts $\hat{N}_j$ and  $\hat{J}_j$ with integer spectrum, which represent symmetric and anti-symmetric combinations of chiral densities \cite{Haldane_1981,sénéchal1999introductionbosonization,Giamarchi2003,Cazalilla_2004,Cazalilla2011}, 
\begin{align}
    &\hat{\phi}_j(x+L)-\hat{\phi}_j(x)=-\pi \hat{N}_j,
    \\
    &\hat{\theta}_j(x+L)-\hat{\theta}_j(x)=\pi \hat{J}_j,\label{zeromode1pbc}
\end{align}
Furthermore, additional restrictions on $\hat{N}_j$ and $\hat{J}_j$ may arise from the identification of low-energy fields with microscopic operators, for example in \refEq{lowenergyürojection}. The quantum numbers are added or subtracted by the action of their dual zero mode operators $\hat{\theta}_{0,i},\hat{\phi}_{0,i}$,
\begin{align}
[\hat{N}_j, \hat{\theta}_{0,i}] = i \delta_{i,j},\quad [\hat{J}_j, \hat{\phi}_{0,i}] = i \delta_{i,j}.\label{zeromode2pbc}
\end{align}
Taking into account \refEq{zeromode1pbc} and \refEq{zeromode2pbc}, we can diagonalize \refEq{Hamiltonian General PBC} by a decomposition of the fields in terms of the bosonic modes $\hat{\beta}_{\sigma},\hat{\beta}_{\tilde{\sigma}}$ and the zero mode operators,
\begin{align}
\left[\hat{\beta}_{n,\sigma},\hat{\beta}^{\dagger}_{\tilde{n},\tilde{\sigma}}\right]=\delta_{n,\tilde{n}}\delta_{\sigma,\tilde{\sigma}},\label{bosonicommutator}
\end{align}
Here $\sigma,\tilde{\sigma}=1,2$, where the bosonic modes describe excitations of the ground state with velocities given by \refEq{eq: soundvelocitiesbulk}.  We find the following finite-size mode decomposition for the fields $\hat{\phi}_j(x,t)$,
\begin{align}
    &\hat{\phi}_j(x,t)
    = \hat{\phi}_{0,j}(x,t)
     + \hat{\phi}_j^{\mathrm{osc}}(x,t),
    \label{eq:PBC_phi_full}
    \\
&\hat{\phi}_{0,j}(x,t)=\hat{\phi}_{0,j}- \frac{\pi \hat{N}_j}{L}\,x+\left(\frac{v_{j} K_{j}\pi }{L }\hat{J}_{j}-\frac{g_{i\neq j}}{4L}\hat{N}_{i\neq j}\right)t\label{eq:PBC_phi_zero},
    \\
   & \hat{\phi}_j^{\mathrm{osc}}(x,t)
=
\sum_{n\neq 0,\sigma=1,2}
\frac{e^{i\frac{2\pi n}{L}(x- \operatorname{sgn}(n)\tilde{v}_\sigma t)}}{\sqrt{2\pi L|n|}}
C^{(n)}_{j,\sigma}\,\hat{\beta}_{\sigma,n}
+ \text{h.c.}
.
    \label{eq:PBC_phi_osc}
  \end{align}
The matrix in \refEq{eq:PBC_phi_osc} obeys the conjugation relation $C^{(-n)}_{j,\sigma}=\bar{C}^{(n)}_{j,\sigma}(n)$ and inherits the algebraic exponents of correlation functions,
   \begin{align}
    & C^{(n)}_{j,\sigma}= \begin{pmatrix}
  \sqrt{
        \frac{\pi v_1 K_1}{2\tilde{v}_1}
        \frac{\tilde{v}_1^2 -\bar{v}_{2}^2}{\tilde{v}_1^2 - \tilde{v}_2^2}
    }, &  \sqrt{
        \frac{\pi v_1 K_1}{2 \tilde{v}_2}
        \frac{\bar{v}_{2}^2 - \tilde{v}_2^2}{\tilde{v}_1^2 - \tilde{v}_2^2}
    }\\ 
  \,\operatorname{sgn}(n)\,\sqrt{
        \frac{\pi v_2 K_2}{2 \tilde{v}_1}
        \frac{\tilde{v}_1^2 - \bar{v}_{1}^2}{\tilde{v}_1^2 - \tilde{v}_2^2}
    } & \,\operatorname{sgn}(n)\,\sqrt{
        \frac{\pi v_2 K_2}{2 \tilde{v}_2}
        \frac{\bar{v}_{1}^2 - \tilde{v}_2^2}{\tilde{v}_1^2 - \tilde{v}_2^2}
   }
\end{pmatrix}\label{matrix}
\end{align}
It should be noted that the off-diagonal entries of $C^{(-n)}_{j,\sigma}$ vanish exactly at the fine-tuning, see \refEq{decoupledLL0}. 
The momenta conjugated to the density fields are from
\begin{align}
&\hat{\Pi}_j(x,t)=\frac{1}{\pi}\partial_x \hat{\theta}_j (x,t)\nonumber
    \\
&\quad\quad\quad\phantom{a}
    = \frac{1}{\pi v_j K_j}
    \left(
        \partial_t \hat{\phi}_j(x,t)
        - \frac{g_{i \neq j}}{4\pi}\,\partial_x \hat{\phi}_{i \neq j}(x,t)
    \right),
    \label{eq:PBC_Pi2_deriv}
    \\
    &\hat{\theta}_{0,j}(x,t)=\hat{\theta}_{0,j}+\frac{\pi \hat{J}_j}{L}x+\left(\frac{v_j \pi}{K_jL}\hat{N}_j-\frac{g_j}{4L} \hat{J}_{i\neq j}\right)t,
\end{align}
which couples excitations as well as the topological sectors of the two chains.
The diagonalized form splits into an $L$-dependent zero mode Hamiltonain $\hat{H}_0$ and excitations with energy density $\hat{h}^{PBC}_{\sigma,n}$. The excitations correspond to the eigenmodes of the infinite system in \refEq{Hamiltonian General PBC}, after appropriate continuization of the discrete momenta $k_n$, and the zero-modes remain coupled by the interactions of both chains.
\begin{align}
k_n=&\frac{2\pi \vert n \vert}{L},
\\
    \hat{H} =& \sum_{n\neq0,\sigma=1,2} \hat{h}^{PBC}_{\sigma,n}+\frac{1}{L}\hat{H}_0,\label{eq: Diagonalhamiltonianpbc}
\end{align}
where,
\begin{align}
   \hat{h}^{PBC}_{\sigma,n}=&k_n \tilde{v}_\sigma \left(\hat{\beta}^{\dagger}_{n,\sigma}\hat{\beta}_{n,\sigma} +\frac{1}{2}\right),\label{eq: excitationhamiltonianpbc}
\\
    \hat{H}_0 =&\sum_{j=1}^2\frac{v_j\pi}{2} \left( \frac{{\hat{N}_j}^2}{K_j} + K_j\hat{J}_j^2 \right )  - \sum_{i\neq j}\frac{g_i}{4 } \hat{N}_i \hat{J}_{j}. \label{eq: zeromodehamiltonianpbc}
\end{align}
The velocities of the excitations $\tilde{v}_{1,2}$, as defined in \refEq{eq: soundvelocitiesbulk}, parametrically depend on the inter-chain coupling and so dynamical instabilities can develop.
The following inequalities are found, corresponding to the trace and determinant instabilities of the eigenmodes in \refEq{trace0bulk} and \refEq{determinant0bulk} respectively, where positive semi-definiteness according to \refEq{stabilityfluxtube} is viewed as superordinate,
 \begin{align}
    &q_1q_2\geq -\frac{1}{2}\left(s+\frac{1}{s}\right) ,\quad s=\frac{v_1}{v_2} \label{eq: trace}
    \\
    & (q^2_1-1) (q^2_2-1)\geq0.\label{eq: determinant}
\end{align}
\refEq{determinant0bulk} is always fulfilled under the bound in \refEq{stabilityfluxtube}, where its saturation point implies 
that the slow-channel $\tilde{v}_2$ becomes exactly zero. Similarly, \refEq{eq: trace} is fulfilled almost everywhere, and saturation is only possible if we additionally have the following, 
\begin{align}
    &s=1, \phantom{a}q_1=\pm1,\phantom{a}q_2=\mp1,
\end{align}
where both channels become dynamically unstable and the Luttinger description breaks down per se \cite{Giamarchi2003,Cazalilla_2004,Cazalilla2011}. 
The fine-tuning in \refEq{Dirichletboundaryissue3} reduces the critical line, which represents the partial instability, to a single point. In this case, the the slow mode becomes unstable first, for generic bare sound velocities $v_{1,2}$,
as the whole system becomes unstable already at $\tilde{v}_2\rightarrow 0$. However, exactly when the bare sound velocities are equal, i.e., $v_1=v_2$, both channels disperse equally and thus become critical at the same time. 
\\
\\
Positive semi-definite sound-velocities, i.e., $\tilde{v}_{1,2}\geq 0$, implies that the ground-state is represented by the zero mode Hamiltonian $\hat{H}_0$ in \refEq{eq: zeromodehamiltonianpbc}.
The set of zero modes for number $(\hat{N}_1,\hat{N}_2)$ and current $(\hat{J}_1,\hat{J}_2)$ excitations is generally restricted to $\mathbb{Z}$, where particular constraints on these integers, i.e., the so-called selection rules, are derived only from imposing boundary conditions on the low-energy projection of a physical operator \cite{Haldane_1981,sénéchal1999introductionbosonization}. 
\\
\\
The zero mode Hamiltonian in \refEq{eq: zeromodehamiltonianpbc} can always be diagonalized by a linear transformation, under the requirement of \refEq{stabilityfluxtube}. However, only \emph{modular} transformations map grids of integers to dual ones, and thus preserve the integrality of quantum numbers and their selection rules \cite{Pletyukhov2004}. Otherwise, the entire set $(\tilde{N}_1,\tilde{J}_1,\tilde{N}_2,\tilde{J}_2)$ must be  mutually considered in minimizing $\hat{H}_0$, and the sectors remain intertwined. A possible choice for such a \emph{modular} transformation is the following,
\begin{align}
&\tilde{N}_{1,2}=\hat{N}_{1,2},\label{eq: modular}
\\
    &\tilde{J}_1=\hat{J}_1-\frac{\alpha_2}{\pi}\hat{N}_2,\nonumber
    \\
     &\tilde{J}_2=\hat{J}_2-\frac{\alpha_1}{\pi}\hat{N}_1,\nonumber
     \\
     &\tilde{H}_0 =\sum_{j=1}^2\frac{ \pi}{2L} \left( \frac{\tilde{v}_{j}}{K_j}{\tilde{N}_j}^2 +v_j K_j\tilde{J}^2_j \right )\label{eq: modularHamiltonian},
\end{align}
where $\frac{\alpha_{1,2}}{\pi}\in\mathbb{Z}$. The ground state energy is subsequently found by minimizing the two channels $(\tilde{N}_1,\tilde{J}_1$) and ($\tilde{N}_2,\tilde{J}_2)$ independently.
A restriction of the scaled flux-tube constants $\frac{\alpha_{2,1}}{\pi} $ to $\mathbb{Z}$ is generally not possible, and thus the zero mode sectors of the two chains are dynamically coupled by the Hamiltonian. From the fine-tuning in \refEq{Dirichletboundaryissue3} we can conclude that the anyonic fixed point never admits decoupled zero modes besides the trivial, bosonic and fermionic, limits $\frac{\alpha}{\pi}=0,1$.
Finally, even if such modular transformations exist, the spectral properties of the constituents of $\hat{H}_0$ are still intertwined, in analogy to the incomplete left-right, or spin-charge, separation for Luttinger liquids on a ring \cite{sénéchal1999introductionbosonization}. That is, the selection rules dictate the allowed parities of the integers $(\hat{N}_1,\hat{J}_1$) and ($\hat{N}_2,\hat{J}_2)$, so the integer lattice spanned by these quantum numbers is not a simple cartesian product.

\subsection{Open Boundary Conditions}
\label{subsec: openboundaryconditions}

As we have seen from the semi-infinite system in \refEq{eq: hamiltoniandensity}, our model is only conformally invariant in the presence of a boundary under the fine-tuning conditions in \refEq{Dirichletboundaryissue3} \cite{Recknagel_Schomerus_2013}. In the case of actual Dirichlet boundary conditions, vertex operators have to vanish at two boundaries $x=0,L$. As the fields are compactified, this implies that the density field is defined up to the operators $\hat{N}_j$ with the spectrum in $\mathbb{Z}$,
\begin{align}
    &\hat{\phi}_j(L)-\hat{\phi}_j(0)=\gamma_j+\pi \hat{N}_j,\label{zeromode1obcbc}
\end{align}
In contrast to periodic boundary conditions, the endpoints of the fields are not identified but are fixed independently at $x=0,L$, where $\gamma_j$ corresponds exactly to their difference. The exact value of $\gamma_j$ is fixed by the explicit low-energy projection of microscopic operators, where further restrictions on $\hat{N}_j$ can arise from selection rules. The phase fields $\hat{\theta}_{j}$ do not obey constraints analogous to \refEq{zeromode1obcbc} therefore there exists only one pair of zero modes, 
\begin{align}
        [\hat{N}_i, \hat{\theta}_{0,j}] = i \delta_{i,j},\label{zeromodeobccommutator}
\end{align}
This reflects the fact that zero-mode current fluctuations are absent in the Dirichlet case, since only one chiral species exists that is folded at the boundaries \cite{Giamarchi2003,Cazalilla_2004,Recknagel_Schomerus_2013}. Conversely, there cannot exist a dual operator analogous to \refEq{zeromodeobccommutator} and so the zero mode of the density field  is not operator-valued, i.e., $\hat{\phi}_{0,j}\equiv\phi_{0,j}$, but just a constant that fixes the Dirichlet boundary conditions uniquely up to compactification. The finite mode components, on the other hand, are again represented by ordinary bosons in analogy to \refEq{bosonicommutator}.
\\
\\
The fine-tuning condition in \refEq{Dirichletboundaryissue3} leads to a manifestly d'Alembertian form for the equations of motion corresponding to \refEq{Hamiltonian General PBC} \cite{Alembert1749},
we can therefore use a sinusoidal mode decomposition for the $\hat{\phi}$ fields \cite{Eggert92,Gogolin1995,Eggert1997},
\begin{align}
    &\hat{\phi}_j(x,t)=\hat{\phi}_{j,0}(x) +\hat{\phi}^{\mathrm{osc}}_j(x,t),\label{Dirichletphi}
    \\
    &\hat{\phi}_{j,0}(x)=\phi_{0,j} - (\gamma_j +\hat{N}_j \pi) \frac{x}{L},
    \\
     &\hat{\phi}^{\mathrm{osc}}_j(x,t)=\sum_{n>0}\sqrt{\frac{ v_j K_j}{ \tilde{v}_j n  }} \sin\left(\frac{\pi n}{L}x\right)\left(\hat{\beta}^{\dagger}_{j,n}e^{i \bar{v}_j\frac{\pi n}{L}t }+\mathrm{h.c.}\right).
\end{align}
Here, the expansion coefficients of the standing waves correspond to the diagonal entries of the periodic matrix $C^{(n)}_{j,\sigma}$ after inserting \refEq{Dirichletboundaryissue3} and a suitable renormalization.
The canonical momenta, on the other hand, are again modified analogously to the flux-tube attachment in \refEq{fluxtubeattachement},
\begin{align}
   &\hat{\Pi}_j(x,t)=\frac{1}{\pi}\partial_x \hat{\theta}_j (x,t)\nonumber
   \\
     &\phantom{a} \quad\quad\quad= \frac{1}{\pi  K_j v_j}
  \left(
        \partial_t \hat{\phi}_j(x,t)
        - \frac{g_{i\neq j}}{4\pi}\,\partial_x \hat{\phi}_{i\neq j}(x,t)
    \right),\label{Dirichletpi}
    \\
    &    \hat{\theta}_{j,0}(x,t) = \hat{\theta}_{0,j} - \frac{\bar{v}_j}{K_j L}\left(\gamma_j  +\hat{N}_j \pi\right) t   \nonumber
    \\
&\quad\quad\quad\phantom{a}+\frac{1}{v_j K_j} \frac{g_{i\neq j}}{4\pi}  \frac{\gamma_{i \neq j}  +\hat{N}_{i\neq j} \pi}{L} x\label{dirichletthetazeromode},
\end{align}
where we fix the integration constant in \refEq{Dirichletpi} so that $\hat{\theta}^{\mathrm{osc}}_j(x,t)$ does not have a constant contribution, i.e., a $x$-independent term. According to \refEq{dirichletthetazeromode} the linear $x$-dependence of the zero mode $\hat{\theta}_{j,0}(x,t)$ acts effectively as a gauge field for the other chain. 
The diagonalized form of the Hamiltonian decomposes again into zero modes $\hat{h}_{0,j}$ and excitations $ \hat{h}^{OBC}_{j,n}$, which  a linear dispersion and standing-wave momenta $k_n$, with
\begin{align}
k_n&=\frac{\pi n}{L},
\\
    \hat{H} &=\sum_{j=1,2}\Bigg[ \sum_{n>0}  \hat{h}^{OBC}_{j,n} + \frac{1}{L}\hat{h}_{0,j}\Bigg],
\end{align}
and
\begin{align}
\hat{h}^{OBC}_{j,n}&=\frac{\pi n}{L} \tilde{v}_j\left(\hat{\beta}^{\dagger}_{n,j}\hat{\beta}_{n,j} +\frac{1}{2}\right)\label{excitationsmodeopen},
\\
    \hat{h}_{0,j}&=\frac{\tilde{v}_j}{2\pi K_j }  \bigg( \gamma_j +\hat{N}_j \pi\bigg)^2.\label{zeromodeopen}
\end{align}
In contrast to the periodic case in \refEq{eq: zeromodehamiltonianpbc}, the zero mode Hamiltonian completely decouples. Furthermore, the summation over the finite mode components reflects the presence of only a single chiral species at the level of excitations. 
Furthermore, we can conclude from \refEq{excitationsmodeopen} and \refEq{zeromodeopen} that the zero modes have exactly the same stability requirement as the excitations given by \refEq{decoupledLL0}.

\section{Correlation Functions}
\label{sec: correlationfunctions}

After having completed the diagonalization of the Hamiltonian in \refEq{Hamiltonian General PBC} for cyclic and Dirichlet boundary conditions, we apply it to the spin zigzag chain realization of Ref. \cite{Batista2014}. In that case, the universal parameters are given by \refEq{kbatista} to \refEq{gbatista}, where the low-energy projection of spin operators follows from \refEq{lowenergyürojection},
\begin{align}
    \hat{S}^z_x&=\sum_{\sigma=1,2}\left(\frac{1}{\pi}\partial_x\hat{\phi}_\sigma(x)+A\cos(2\hat{\phi}_\sigma(x))\right)+\cdot\cdot\cdot,\label{lowenergylongitudinaloperator}
    \\
        \hat{S}^-_x&=\sum_{\sigma=1,2}Be^{i(-1)^{\sigma+1}Qx}e^{-i\hat{\theta}_{\sigma}(x)}e^{i\hat{\phi}_{\tilde{\sigma}\neq\sigma}(x)}+\cdot\cdot\cdot,\label{lowenergyladderoperator}
\end{align}
where $A$ and $B$ denote non-universal constants suubleading corrections are indicated.
For this realization, we discuss two characteristic signatures of finite-size Luttinger liquids, persistent currents \cite{Loss1992,Schmeltzer1993,Pletyukhov2004,Minguzzi2025}, and Friedel oscillations \cite{Friedel1958,Gogolin1995,Grabert1995,Eggert2000,Eggert2003}. 
While persistent currents originate from the response of the zero mode with respect to an external gauge flux, the Friedel oscillation is the minimal correlation function that reflects broken translational invariance via an in-homogeneous local density of fermionic excitations. We find parity effects of persistent currents, similar to ordinary fermionic Luttinger liquids, that are intertwined with commensurability effects due to special values of the statistical parameter $\alpha$. Likewise, statistical transmutation is probed by pronounced Friedel oscillations, where we also find parity effects with respect to the number of excitations.
For consistent universal parameters according to \refEq{Dirichletboundaryissue3}, we  additionally calculate the longitudinal spin-correlations for both types of boundary conditions.  
In the finely tuned case the two channels decouple and we recover the usual bulk and boundary distinction of correlation functions known from conformal field theory \cite{Cardy1984,cardy1991a}. We find chirally imbalanced exponents in subleading contributions, which can be probed by spectroscopic measures. However, correlation functions remain parity symmetric as the underlying field theory at fine tuning \cite{Batista2014}.   

\subsection{Persistent Currents}
\label{subsec: persistentcurrents}
At first, we investigate a representative zero mode effect of cyclic Luttinger liquids, persistent currents \cite{Loss1992,Schmeltzer1993,Pletyukhov2004,Minguzzi2025}.
Selection rules are generally attached to the exchange statistics of the physical entities that are considered, as well as their explicit low-energy projection and boundary conditions \cite{Haldane1981,sénéchal1999introductionbosonization,Cazalilla_2004}, i.e., \refEq{lowenergylongitudinaloperator} and \refEq{lowenergyladderoperator} for Ref. \cite{Batista2014}.
A boundary magnetic flux $\phi\in (0,2]$ that induces the persistent current is considered, whereas the helical pitch angle $Q$ in \refEq{helicalangle} becomes quantized in the cyclic case \cite{Batista2009,Batista2012,Batista2014},
\begin{align}
&\hat{S}^-_{j+L} = \hat{S}^-_{j}e^{i\pi \phi},\label{boundaryphase}
     \\
     &Q=\frac{2\pi n}{L},\label{helicalangleperiodic}
\end{align}
with $n\in[1,L]$.
From \refEq{lowenergyladderoperator} as well as \refEq{boundaryphase} and \refEq{helicalangleperiodic}
we conclude the following selection rules,
\begin{align}
 \hat{N}_1 &=   n_1 + m_1, \\
    \hat{J}_1 &= 2n + \phi + n_2 - m_2 , \\
   \hat{N}_2 &=  n_2 + m_2, \\
    \hat{J}_2 &=-2n + \phi + n_1 - m_1 ,\label{zeromodebatista}
\end{align}
where the spectrum of $n_1, m_1, n_2, m_2$ is in $ \mathbb{Z}$. Interestingly, we find that the constraints on the parity of operators are inter-chain rather than intra-chain, contrary to the low-energy projections of fermionic species \cite{Loss1992,Schmeltzer1993,Pletyukhov2004}.
Ref. \cite{Batista2014} effectively considers the canonical ensemble, as the spin chain is supposed to be close to saturation with a dilute set of fermionic excitation around the lowest weight state. Thus, we fix the magnetization sector of the spin chain  $S^z_T=N-\frac{L}{2}$, by the number of excitations in the two fermionic species $N=\hat{N}_1+ \hat{N}_2$, where we eliminate $\hat{N}_2=N-\hat{N}_1$, and obtain simplified rules,
\begin{align}
 \hat{J}_1 &= \phi + N-N_1+2l ,
 \\
 \hat{J}_2 &=  \phi +N_1+2k,\label{selectionrulebatista2}
\end{align}
with parity-adapted current eigenvalues $l,k\in \mathbb{Z}$ that are shifted relative to the energetically irrelevant, helical winding number $\pm 2n$. Consequently, we can express the zero mode energy as follows,
\begin{align}
E^{(0)}_{N_1,k,l}&=E(N_1)+E(N_1,k,l),\label{zeromodehamiltonianbatista}
    \\
    E(N_1)&=\frac{\pi v}{2L }\left[N^2_1+ (N-N_1)^2\right]\left(1-\left(\frac{\alpha}{\pi}\right)^2\right),\label{chargingpartzeromode}
    \\
    E(N_1,k,l)&=\frac{2\pi v}{L}\left[l+\Delta_2(N_1)\right]^2+\frac{2\pi v}{L}\left[k+\Delta_1(N_1)\right]^2,\label{currentpartzeromode}
    \\
    \Delta_1(N_1)&=\frac{1}{2}\left[\left(1+\frac{\alpha}{\pi}\right)N_1+\phi\right],\label{Delta1}
    \\
        \Delta_2(N_1)&=\frac{1}{2}\left[\left(1-\frac{\alpha}{\pi}\right)\left(N-N_1\right)+\phi\right],\label{Delta2}
\end{align}
where the centers of the parabolas $E(N_1,k,l)$ spanned by $k$ and $l$, are in-homogeneously shifted by the statistical parameter $\alpha$ and the density zero mode $N_1$. Furthermore, we can express the stability criterion of the Luttinger Hamiltonian according to \refEq{stabilityfluxtube}, purely in terms of the statistical parameter, 
\begin{align}
   \alpha\leq \pi,
\end{align}
where stability is ensured for the full range of statistical transmutation. 
\\
\\
As discussed in section \refSec{subsec: Diagonalizationperiodic} the zero mode Hamiltonian of Ref. \cite{Batista2014}, is only modular invariant in the non-anyonic cases of bosons and fermions. Hence, the partition function $Z(\beta,\phi,N)$ generally does not decouple into disjoint spectral sectors \cite{Pletyukhov2004}. Here, $\beta$ denotes the nondimensional inverse temperature with $k_B=1=\hbar$. Likewise, $L$ denotes the number of lattice sites, not the physical length. Mathematically, the coupling of spectral sectors implies that the zero-mode partition function corresponds to Siegel/ generalized $\Theta$ functions \cite{Abramowitz1964,Pletyukhov2004}, rather than an ordinary Jacobi $\Theta$ function, i.e.,
\begin{align}
    &\Theta^{(x)}\left(a,b\right)=\begin{cases}
             \Theta_3(a,b) \phantom{a}x\phantom{a}\mathrm{even},\\
            \Theta_2(a,b) \phantom{a}x\phantom{a}\mathrm{odd},
           \end{cases}
            \\
           & \Theta_2(a,b)=\sum_{n\in\mathbb{Z}}b^{(n+1/2)^2}e^{(2n+1)\pi i a },
           \\
           & \Theta_3(a,b)=\sum_{n\in\mathbb{Z}}b^{n^2}e^{2\pi i a n},
\end{align}
where $b=e^{i\pi \tau}$
Nevertheless, it is instructive to expand the partition function $Z(\beta,\phi,N)$ in a double Fourier series that involves these functions, namely a series of flux momenta $k$ conjugated to $\phi$, as $Z(\beta,\phi+2)=Z(\beta,\phi)$, and of a dual variable $\zeta_k$ conjugated to $\hat{N}_1$. By using Poisson resummation we find the following representation for the zero mode partition function, 
\begin{align}
    &Z(\beta,\phi,N)=Z_0(\beta,N)+\sum_{k>0} Z_k(\beta,N)\cos\left(\pi k \phi\right),\label{partitionfunctionbatista}
    \\
           &Z_k(\beta)=\frac{L\tilde{Q}^{\frac{N^2}{2}} }{2\beta v}\tilde{q}^{\frac{k^2}{2}} \Theta^{(k)}\left(\pi \chi_N,\tilde{q}^2\right)\Theta^{(N)}\left(\pi \zeta_k,\tilde{Q}^2\right),\label{partitionfunctiondoublefourier}
           \\
           &\chi_N=\big(1-\alpha/\pi\big)N/2,\quad \zeta_k=-\big(1-\alpha/\pi\big)k/2,
           \\
           &\tilde{Q}=e^{-\frac{\beta\pi v}{2L}\big(1-\left(\alpha/\pi\right)^2\big)},\quad \tilde{q}=e^{-\frac{L \pi}{2\beta v}}.
\end{align}
The effect of the statistical parameter $\alpha$ on \refEq{partitionfunctiondoublefourier} is two-fold. 
At first, $\alpha$ mediates interactions via $\tilde{Q}$, i.e., by reducing the density stiffness for increasing statistical interactions \cite{Loss1992,Schmeltzer1993}. Consequently, this increases the relevance of the current-density terms in \refEq{currentpartzeromode} in the branch competition with \refEq{chargingpartzeromode}.
Secondly, $\alpha$ alters the arguments of the $\Theta$ functions $\chi_N$ and $\zeta_k$, and so manifests statistical transmutation by changing the dominant zero mode contribution.
\\
\\
Generalized $\Theta$ functions are more difficult to analyze in contrast to ordinary ones \cite{Pletyukhov2004}, but \refEq{partitionfunctionbatista} can be severely simplified if the statistical parameter is a rational multiple of $\pi$, i.e.,
\begin{align}
\alpha/\pi=p/q,\quad \mathrm{gcd}(p,q)=1.
\end{align}
In  this case, the infinite series over flux momenta $k$ can be reduced to a finite number of residue classes, i.e.,
\begin{align}
&M=\frac{2q}{\mathrm{gcd}(q\pm p,2q)},\label{definitionofm}
\\
   & k=r+Pn, \label{restclass}
   \\
   &P=\mathrm{lcm}\left(M,2\right),
\end{align}
with $ n\in \mathbb{Z}$ and $r\in[0,P-1]$. In this case, the partition function is simplified in terms of ordinary $\Theta$ functions and those with rational characteristic $C_r(\phi)$ \cite{Abramowitz1964},
\begin{align}
 &Z(\beta,\phi,N)\overbrace{=}^{\frac{\alpha}{\pi}=\frac{p}{q}\in\mathbb{Q}}\mathcal{Z}(\beta,\phi,N)=\frac{L\tilde{Q}^{\frac{N^2}{2}} }{2\beta v}\sum_{r=0}^{P-1}Z_r(\beta,N,\phi),\label{finiteresiduepartition}
 \\
& Z_r(\beta,N,\phi)=\Theta^{(r)}\left(\frac{\pi m}{M}N,\tilde{q}^2\right)\Theta^{(N)}\left(-\frac{\pi m}{M}r,\tilde{Q}^2\right)C_r(\phi),
\\
   &C_r(\phi)=\sum_{n \in\mathbb{Z}} \tilde{q}^{(r+nP)^2/2}e^{i\pi (r+nP)\phi},
   \\
          & m=\frac{q-p}{\mathrm{gcd}(q-p,2q)}.
\end{align}
The ordinary $\Theta$ functions are even and odd with respect to a $\pi$ shift of their first arguments, which implies the following relations for the persistent current,
\begin{align}
&F=-\frac{1}{\beta} \ln\left(\mathcal{Z}(\beta,\phi,N)\right)\label{freeenergydefinition}
\\
&\mathcal{I}(\phi,\beta,N)=-\frac{\partial F}{\partial \phi},\label{pesistentcurrentdefinition}
\\
    &\mathcal{I}(\phi,\beta,N+M)\overbrace{=}^{M\phantom{a}\mathrm{even}}\mathcal{I}(\phi+1,\beta,N),\label{persistentcurrentphishiftevenM}
    \\
    &\mathcal{I}(\phi,\beta,N)=\mathcal{I}(\phi,\beta,N+2M),\label{persistentcurrentperiodicity}
    \\
    &\mathcal{I}(\phi,\beta,N)=\mathcal{I}(\phi,\beta,2M-N).\label{persistentcurrentreflection}
\end{align}
Furthermore, we conclude from \refEq{finiteresiduepartition}, that the persistent current shows period-halving in the flux-variable $\phi$, and in the number of excitations, for special $N$ and even $M$,
\begin{align}
      &mN\equiv \frac{M}{2}\quad (\mathrm{mod}\phantom{a} M),\phantom{a} M\phantom{a}\mathrm{even}\label{periodhalvingcurrent}
      \\
      &\Longrightarrow \mathcal{I}(\beta,\phi,N+M)=\mathcal{I}(\beta,\phi,N),
    \\
    &\Longrightarrow \mathcal{I}(\beta,\phi+1,N)=\mathcal{I}(\beta,\phi,N),
\end{align}
due to the dependence of \refEq{finiteresiduepartition} on the parity of $N$.
\refEq{periodhalvingcurrent} is not the only number-theoretic effect. If $N$ corresponds to an integer multiple of $M$, the current-density couplings of the zero modes in \refEq{Delta1} and \refEq{Delta2} become equal modulo an integer, and the two channels are fully in phase,
\begin{align}
  &N\equiv 0\quad(\mathrm{mod}\phantom{a} M)
    \\
    &\Longrightarrow \Delta_1(N_1)\equiv \Delta_2(N_2)\quad  (\mathrm{mod}\phantom{a} 1)\label{singlesawtoothcondition}.
\end{align}
\\
\\
Even-odd effects as in \refEq{partitionfunctiondoublefourier}, or \refEq{periodhalvingcurrent} and \refEq{singlesawtoothcondition}, are best illustrated at zero temperature, so we investigate this regime in more detail. In this case, the persistent current is given in closed form as a sum of two sawtooh functions $s(x)$, one for each channel,
\begin{align}
    &I_{T=0}=-\frac{\pi v}{2L}\left[s(\Delta_1(\bar{N}_1))+s(\Delta_2(\bar{N}_1))\right]\label{sawtooth},
    \\
    &s(x)=2\left(x-\lfloor x+1/2 \rfloor\right), \nonumber
\end{align}
where $s(x)$ is undefined at $x-1/2\in\mathbb{Z}$.
$\bar{N}_1$ in \refEq{sawtooth} corresponds to the argument that minimizes the ground state energy in \refEq{zeromodehamiltonianbatista}, where the inextricable coupling of zero mode sectors discussed in section \refSec{subsec: Diagonalizationperiodic} complicates this endeavor. However, if we minimize first the current quantum numbers, the zero-mode energy that couples to $\phi$ depends on $N_1$ solely via the rest class $r$, i.e., $N_1(r)=r+n M$. Subsequently, the minimization of the charging energy $E(N_1)$ in \refEq{chargingpartzeromode} fixes the index $n$, for a constant rest class $r$, while the optimal $r^*$ has to be found consecutively, 
\begin{align}
    &E^{(0)}_{r}\equiv E(N_1(r))+\frac{\pi v}{L}\left[s(\Delta_1(r))^2+s(\Delta_2(r))^2\right]\label{zerotemperatureenergyperiodic},
\\
    &\bar{N}_1=N_1(r^*),\quad N_1(r)=r+M\cdot\mathrm{nint}\bigg(\frac{N/2-r}{M}\bigg),\label{minimalN1}
    \\
    &r^*=\arg \min_{r\in[0,M-1]} E^{(0)}_{r}.\label{minimalbranch}
\end{align} 
Insertion of the minimum condition in \refEq{minimalN1} into \refEq{sawtooth} gives the current per branch,
\begin{align}
    I^{(r)}_{T=0}(\tilde{\phi}_r)&=-\frac{\pi v}{L}\left[s\bigg(\tilde{\phi}_r\bigg)+s\bigg(\tilde{\phi}_r+\delta_N\bigg)\right],\phantom{a}  0\leq \delta_N\leq 1/2\nonumber
    \\
   & =-\frac{\pi v}{L}\begin{cases}
           4\bar{\phi}_r+2\delta_N \phantom{abcde}-\frac{1}{2}<\bar{\phi}_r<\frac{1}{2}-\delta_N,\\
         4\bar{\phi}_r+2\delta_N-2\phantom{abcde} \frac{1}{2}-\delta_N<\bar{\phi}_r<\frac{1}{2},
        \end{cases}\label{sawtoothbranch}
\end{align}
with
\begin{align}
    \tilde{\phi}_r=&\frac{\phi}{2}-\frac{m r}{M},\quad \delta_N=\frac{m N}{M}\phantom{a}\mathrm{mod}\phantom{a} 1,
\end{align}
where the actual persistent current is the piecewise selection of branch currents with optimal $r^*$.
\refEq{sawtoothbranch} reveals a mutual flux dependence $\tilde{\phi}_r$ per branch, while one of the currents is horizontally shifted by $\delta_N$. From the branch current, amplitude doubling of the current according to \refEq{singlesawtoothcondition} becomes manifest as $\delta_N\equiv0$, whereas the period-halving in \refEq{periodhalvingcurrent} follows from $\delta_N\equiv1/2$. The shift $\delta_N$ equips the branch current with a larger arithmetic structure compared to ordinary Luttinger liquids \cite{Loss1992,Pletyukhov2004}, i.e.,  $N \phantom{a}\mathrm{mod} \phantom{a}M$, which is extended to $N \phantom{a}\mathrm{mod} \phantom{a}2M$ by the parity dependence of the charging energy. 
\\
\\
In the following, we discuss the properties of the discontinuities of the current as a function of the flux and classify different types of transition between zero mode sectors. With this we distinguish ordinary parity effects of fermionic Luttinger liquids with commensurability effects induced by statistical transmutation.
The energy in \refEq{zerotemperatureenergyperiodic} is continuous but shows cusps at specific flux values $\phi_c$. Consequently, these cusps correspond to discontinuities of the persistent current, which can appear either at symmetric or non-symmetric flux values,
\begin{align}
    \begin{cases}
          \phi_c= \phi^{s}_c \quad \phi^{s}_c\equiv - \phi^{s}_c\phantom{a}\mathrm{mod}\phantom{a} T, \\
        \phi_c\neq \phi^{s}_c,\phantom{a} 
        \end{cases}\label{fluxpositions}
\end{align}
with $T$ being the respective $\phi$-periodicity of the energy. The cusps at $\phi^s_c$ are generally mirror symmetric around the critical flux, while the cusps at other flux values are generically tilted but can become symmetric at fine-tuning. 
As the flux $\phi$ is the only external parameter, the cusps appear generically inter-branch or by the degeneracy of only two residuals, i.e., $E^{(0)}_r=E^{(0)}_{\tilde{r}}$. However, also multi-coincidences may appear where three or more residue classes become iso-energetic for some parameters. 
\\
\\
The generic 
cusps can be distinguished by their left- and right-sided limits of the minimizing branch,
\begin{align}
    r_{\pm}=\lim_{\epsilon\rightarrow 0}  r^{*}(\phi_c \pm\epsilon).
\end{align}
The type $\mathrm{A}$ cusp exists whenever $r_-=r_+$ holds, for flux values $\phi_c$ where one of the branch-sawtooth functions in \refEq{sawtoothbranch} becomes ill-defined.
\\
\\
For $r_-\neq r_+$ the cusps are separable into two other sub-types, whereas contrary to type $A$ the discontinuities of the current arise inter-branch. Type $\mathrm{B}1$ appears exclusively for odd $N$ at symmetric flux values $\phi^{s}_c$, when the limiting branches coincide with the residue class favored by the charging part $E(N_1(r))$, i.e.,
\begin{align}
    r^{(c)}_{\pm}\overbrace{\equiv}^{\mathrm{Mod}[N,2]=1}\frac{N\pm 1}{2} \phantom{a} \mathrm{mod}\phantom{a} M\label{parityeffectbatista}.
\end{align}
This corresponds to the ordinary parity effect of cyclic Luttinger liquids \cite{Loss1992,Schmeltzer1993,Minguzzi2025}, i.e., the degenerate minimal branch in \refEq{parityeffectbatista} implies that the current response is paramagnetic at $\phi_c$, i.e.,
\begin{align}
I_{T=0}(\phi)=\begin{cases}
          I^{(r^{(c)}_-)}_{T=0}(\tilde{\phi}_{r^{(c)}_-})\quad \quad\phi<\phi^{s}_c, \\
        I^{(r^{(c)}_+)}_{T=0}(\tilde{\phi}_{r^{(c)}_+}) \quad\quad\phi>\phi^{s}_c,
        \end{cases}
\end{align}
Type $\mathrm{B}2$ exists whenever $r_-\neq r_+$ is in force, but 
the requirements of type $\mathrm{B}1$ are not strictly obeyed. In particular, this contains different left and right-sided residues at non-symmetric critical flux values $\phi_c\neq\phi^{s}_c$, or minimizing
residues that appear at symmetric flux value $\phi_c=\phi^{s}_c$ that are not equal \refEq{parityeffectbatista}. Besides general multi-residual coincidences, involving three or more energies, there exist other scenarios that exceed the classification into Type $A$ and $B$. For example, if a single or multiple competing branch energies have additionally an internal kink of type $A$ for some $\phi_c$.
\\
\\
\begin{figure*}
    \centering
    \includegraphics[width=0.42\linewidth]{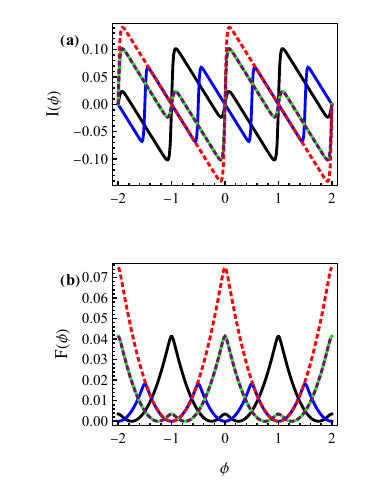}
    \hfill
    \includegraphics[width=0.53\linewidth]{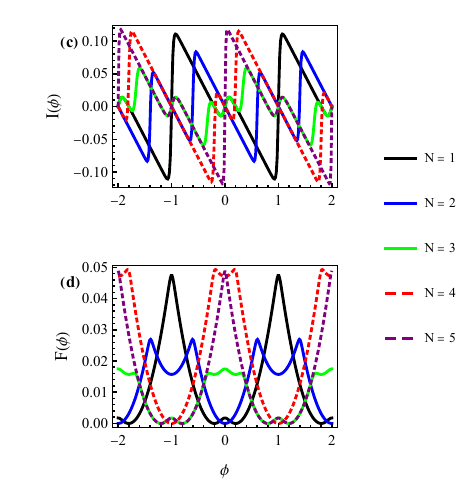}

    \caption{Commensurability and parity effects of persistent current with corresponding free energy for different statistical parameters $\alpha$ for very low temperatures. We consider a dimensionless inverse temperature of $\beta=500$ as well as a unit sound velocity $v=1$ and a number of $L=40$ lattice sites. For rational $\alpha/\pi$, energy and its derivative are repetitive in the total number of excitations $N$ with minimal period $M$. Cusps in the free energy, i.e., $b)$ and $d)$, correspond to actual discontinuities in its derivatives, i.e., $a)$ and $c)$, where the effect of temperature is to smooth out the sharp edges.
    Left: $\alpha=\pi/2$, i.e., $m=1$ and $M=4$.
    Right:  $\alpha=3\pi/5$, i.e., $m=1$ and $M=5$ .
    In the left panel, all cusps in $b)$ appear at symmetric flux values and are mirror symmetric around $\phi^{(s)}_c$.
    For $\alpha=\pi/2$, there is amplitude doubling for $N=4$ and period-halving for $N=2$ in $a)$ according to \refEq{singlesawtoothcondition} and \refEq{periodhalvingcurrent}, respectively.
    In the right panel, some cusps in $c)$ appear at non-symmetric flux values and are therefore not mirror symmetric, see \refEq{fluxpositions}.
    }
    \label{fig:currentcomparison}
\end{figure*}
To illustrate the plethora of different transitions between zero mode sectors and the corresponding jumps of the currents, both  \ref{freeenergydefinition} and \ref{pesistentcurrentdefinition} are shown in \refFig{fig:currentcomparison}, for $\alpha=\frac{\pi}{2}$ (left panel) and $\alpha=\frac{3\pi}{5}$ (right panel) respectively. We consider a unit sound velocity $v=1$, a dimensionless system length $L=40$, as well as a nondimensional inverse temperature of $\beta=500$, i.e., $k_B=1=\hbar$. The temperature is chosen deep in the quantum regime, in order to show the smoothening of discontinuities of \refEq{sawtooth} while validating a numerical restriction of \refEq{partitionfunctiondoublefourier} to just a few flux momenta $k$. Due to \refEq{persistentcurrentperiodicity} and \refEq{persistentcurrentreflection}, it suffices to investigate excitations up to the respective $M$ for rational values of $\alpha/\pi$, defined in \refEq{definitionofm}. In \refFig{fig:currentcomparison} $a)$ in the left panel, we see that the current for a single excitation (blue) jumps at flux values of $\phi_c=0,\pm 1$, with corresponding cusp-like structure in $b)$. Compared to the ordinary parity effect in Luttinger liquids, i.e., type $B1$, at $\phi_c=0$, these transitions show additional internal kinks in one branch at $\phi_c=\pm 1$, i.e., a hybrid of Type $B1$ and Type $A$. For $N=3$, i.e.,  red-dashed graph, the origins of the jumps are exactly reversed  compared to the $N=1$ case, with a Type $B1$ transition at $\phi_c=\pm 1$ and additional internal kinks at $\phi_c=0$.  
For $N=2$ on the other hand, i.e., blue line, the current's period is halved with jumps at flux values $\phi_c=\pm 1/2,\pm 3/2$, see \refEq{periodhalvingcurrent}. Here, the same branch remains minimal on both sites of the transition, i.e., corresponding to Type $A$. The current for $N=4$ excitations (purple-dashed), has a doubled amplitude with jumps at $\phi_c=0$, see also \refEq{singlesawtoothcondition}, where three residuals become iso-energetic but host a unique minimizer in the vicinity of $\phi_c$. For this choice of statistical parameter and particle numbers, all cusps are symmetric around $\phi_c$. 
\\
\\
For a single excitation (black) the classification of smoothened cusps in $b)$ is analogous to the $\alpha=\pi/2$ case, with a Type $B1$ transition at $\phi_c=0$ with an additional internal kink at $\phi_c=\pm 1$. In the $N=2$ case (blue), we have transitions of Type $A$ at $\phi_c=\pm 3/5,\pm 7/5$ with a unique minimizing residue for both sides of the transition. In this case, the transition is at a non-symmetric flux point and not mirror symmetric around $\phi_c$.
An analogous behaviour is found for three excitations at $\phi_c=\pm 1/3, \pm 5/3$ (green), where the current shows a type $B2$ transition.
On the other hand, at $\phi_c=0, \pm 1$ the current is of the ordinary $B1$ type. The $N=4$ cusp is again asymmetric (red-dashed), showing intra-branch transitions of Type $A$ at $\phi_c= \pm 1/5, \pm 9/5 $. Finally, the case $N=5$ shows transitions for $\phi_c= 0,\pm 1$ that are both of $B1$ Type.

\subsection{Friedel Oscillations}
\label{subsec: friedeloscillations}
 
From the explicit low-energy projections in \refEq{lowenergylongitudinaloperator}
and \refEq{lowenergyladderoperator},  as well as the boundary conditions of the spin operators $\hat{S}^{\pm}_{j=0,L+1}\equiv 0$, we can infer the following restrictions on the previously arbitrary constants of the mode decomposition,
\begin{align}
    &\phi_{0,j}=\pi/2,
    \\
    &\gamma_j=Q(L+1)=\pi\tilde{n}_{\mathrm{OBC}},\label{Qquantizationdirichlet}
\end{align}
where $\tilde{n}_{\mathrm{OBC}}\in\left[1,L+1\right]$.
In contrast to the cyclic case in \refEq{helicalangleperiodic}, $Q$ is  not quantized a priori to form a helical solution for the Dirichlet case \cite{Batista2009,Batista2012,Batista2014}. However, the momenta of low-energy excitations around the helical solution are quantized according to their boundary conditions, therefore only the nearest integer to $\frac{(L+1)}{\pi}Q$ can be resolved in \refEq{Qquantizationdirichlet}. 
\begin{figure*}[!t]
    \centering

    \includegraphics[width=0.4\textwidth]{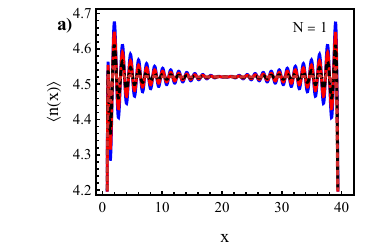}%
    \includegraphics[width=0.52\textwidth]{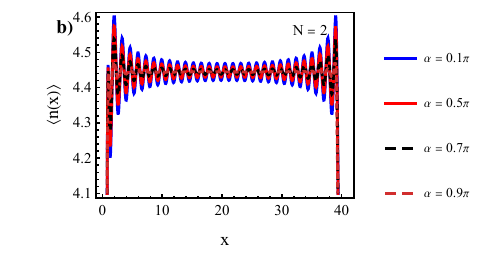}

    \caption{Statistical transmutation revealed by the onset of  density oscillations according to \refEq{Friedelbatista} and parity effects with respect to the total number of excitations $N$, for $L=40$ lattice sites, and $n_{\mathrm{OBC}}=30$, and different values of $\alpha$. Oscillations decrease with increasing $\alpha$, manifesting the absence of Friedel oscillations in the bosonic limit while being pronounced towards the fermionic end \cite{Cazalilla_2004,Eckardt2016,BonkhoffPosske2021}.
    $a$:  Due to an odd number of excitations, i.e., $N=1$, and an even number of lattice sites, the graph has a node between $x_\pm=\frac{L}{2}\pm 1$.
    $b$:  Contrary to the left panel, i.e., $N=2$, there are no additional nodes in the middle of the chain, as both channels in \refEq{Friedelbatista} have exactly the same wave number.
    }
    \label{fig:friedelcomparison}
\end{figure*}
\\
\\
For the model in Ref. \cite{Batista2014}, the local density of fermionic operators according to \refEq{lowenergyürojection} describes fluctuations around the highest-weight state of the saturated spin chain, which oscillate characteristically in the vicinity of boundaries or isolated impurities \cite{Friedel1958,Gogolin1995,Grabert1995,Eggert2000,Eggert2003}. The expansion in \refEq{lowenergyürojection} contains further contributions beyond \refEq{lowenergylongitudinaloperator} or \refEq{lowenergyladderoperator}, but those are eliminated by the neutrality condition which reflects the particle number conservation of chiral species \cite{sénéchal1999introductionbosonization,Giamarchi2003,Cazalilla_2004}. This yields the density oscillation of two gapless modes,
\begin{align}
&\langle\hat{n}(x)\rangle \approx  \frac{\pi }{L+1}\left(2\tilde{n}_{\mathrm{OBC}}-N\right)\nonumber
\\
&- \frac{1}{\pi}\sum_{\sigma=1,2}\left(\frac{\pi \chi}{2(L+1) |\sin(\frac{\pi x}{L+1})|}\right)^{\frac{1}{ \sqrt{1 - \frac{\alpha^2 }{\pi^2}}}}\cos(\epsilon_\sigma x), \label{Friedelbatista}
\\
    &\epsilon_{\sigma}=-  \frac{2\pi }{L+1} \left(\langle \hat{N}_{\sigma}\rangle+\tilde{n}_{\mathrm{OBC}}\right),
\end{align}
shown in \refFig{fig:friedelcomparison} for $L=40$, $N=1,2$, as well as $n_{\mathrm{OBC}}=30$.
Here, $\chi$ corresponds to a short-distance cut-off that is used for the regularization correlation functions \cite{Giamarchi2003}.
The first term in \refEq{Friedelbatista} describes the background density, corresponding to long-wavelength fluctuations of the theory. For the model in reference \cite{Batista2014}, this background corresponds to twice the helical pitch-angle reduced by the number of diluted, magnonic excitations $N$. The second contribution describes oscillations around this background density, which are algebraically decaying from the boundary towards the bulk. 
Due to the fine-tuning in \refEq{kbatista}--\refEq{gbatista} interactions are purely statistical for the model in \cite{Batista2014}, where the decay of each sector is just given by a monotonically decreasing function of the anyonic parameter $\alpha$. In the fermionic limit $\alpha=0$, this decay corresponds to the free fermion fix point $K\equiv 1$ with pronounced oscillations, whereas the bosonic limit $\alpha=\pi$ signals long-range order and thus the breakdown of the theory per se, i.e., $\lim_{\alpha\rightarrow \pi}K\equiv \infty$ \cite{Giamarchi2003,Cazalilla_2004,Cazalilla2011}. The fading of density oscillations away from the fermionic limit $\alpha=0$ is clearly visible. In analogy to Hubbard anyons  \cite{Greschner2014,Eckardt2016,BonkhoffPosske2021,Kwan2024,Kwan2026}, this reveals the full statistical transmutation through a statistically invariant variable \cite{Batista01012004}, i.e., an observable that is the same for anyons or their underlying parent particles.   
\\
\\
The oscillations that decay from the boundary towards the bulk are additionally modulated by a cosine with wavenumber $\epsilon_\sigma$, which depends on the boundary-adapted helical pitch angle $\tilde{n}_{\mathrm{OBC}}$ and on the zero modes \cite{Badalyan2010}. The expectation values of $\langle \hat{N}_j\rangle $ are found by minimizing the zero mode Hamiltonian in \refEq{zeromodeopen}, for the parameter values in \refEq{kbatista}-\refEq{gbatista} and fixed $N=\langle\hat{N}_1+\hat{N}_2\rangle$. We obtain the following solution without the loss of generality
\begin{align}
    &\langle \hat{N}_1 \rangle=\mathrm{nint}\left(\frac{N}{2}\right),\label{minimumzeromodeobc1}
\\
    &\langle \hat{N}_2 \rangle=N-\mathrm{nint}\left(\frac{N}{2}\right),\label{minimumzeromodeobc2}
    \\
    &\operatorname{nint}(x)\equiv
\left\{
n\in\mathbb Z:
|x-n|
=
\min_{m\in\mathbb Z}|x-m|
\right\},
\end{align}
which is independent of $n_{\mathrm{OBC}}$.  \refEq{minimumzeromodeobc1} and \refEq{minimumzeromodeobc2} imply a single wavenumber for the density oscillation, i.e., $\epsilon_{1}=\epsilon_2$, in the case of an even number of excitations as in  \refFig{fig:friedelcomparison} (left). The odd $N$ case is exemplified in \refFig{fig:friedelcomparison} (right) instead, where the Friedel oscillations are additionally modulated by an oscillation with a wavenumber corresponding to the minimal momentum resolution of the lattice, i.e., $\frac{\pi}{L+1}$. Consequently, the oscillatory part has a node that lies at the middle site for odd $L$, or between the adjacent ones for even $L$.

\subsection{Density-Density Correlations}
\label{subsec: densitydensitycorrelation}

In order to investigate both boundary conditions for correlation functions that involve operators at a comparable level, we calculate the density-density expectation value of fermions according to the decomposition of \refEq{batistaderivation}. At fine tuned coupling constants according to \refEq{kbatista}-\refEq{gbatista}, 
this quantity describes longitudinal spin fluctuations of 
Ref. \cite{Batista2014}.
\\
\\
In contrast to other variants of anyonic Luttinger liquids \cite{Greschner2014,BonkhoffPosske2021,BonkhoffEggert2025}, the finely tuned version of \refEq{Hamiltonian General PBC} conserves parity $\hat{P}$, which can be deduced from the transformation behavior of low-energy fields in \refEq{lowenergylongitudinaloperator} and \refEq{lowenergyladderoperator},
\begin{align}
    &\hat{P}\hat{S}^{\pm,z}_{x}\hat{P}^{\dagger}=\hat{S}^{\pm,z}_{L-x+1}\nonumber
    \\
    &\Longrightarrow 
    \begin{cases}
\hat{\phi}_{j}(x)\leftrightarrow -\hat{\phi}_{i\neq j}(-x), \\
            \hat{\theta}_{j}(x)\leftrightarrow \hat{\theta}_{i\neq j}(-x)+(-1)^{j+1}QL,\end{cases}\label{parityonspinoperators}
\end{align}
with $j=1,2$. Spin time reversal on the other hand is broken by an implicit magnetic field that saturates the chain \cite{Batista2014}, whereas the Hamiltonian still preserves anti-unitary conjugation symmetry $\hat{K}$, 
\begin{align}
            & \hat{K}\hat{S}^{\pm,z}_{x}\hat{K}^{\dagger}=\hat{S}^{\pm,z}_{x}\nonumber
    \\
    &\Longrightarrow 
    \begin{cases}
\hat{\phi}_{j}(x)\leftrightarrow \hat{\phi}_{i\neq j}(x), \\
            \hat{\theta}_{j}(x)\leftrightarrow -\hat{\theta}_{i\neq j}(x).\end{cases}\label{complexconjugationonspinoperators}
\end{align}
The transformations in \refEq{parityonspinoperators} and \refEq{complexconjugationonspinoperators} relate excitations at helical wavevectors $\pm Q$ by a permutation of fields. Similarly, correlation functions must be invariant with respect to those transformations. This implies that decay exponents have to appear in appropriate pairs as well, in order to ensure spatio-temporal symmetry of the full correlation function, while the individual chains can still decay chirally asymetric.   
\\
\\
Fermionic densities $\hat{n}_j=\hat{c}^{\dagger}_j\hat{c}_j$ according to \refEq{lowenergyürojection}, describe fluctuations about the highest-weight state of the spin $\hat{S}^x_j$ operator with low-energy projection defined in \refEq{lowenergylongitudinaloperator} \cite{Batista2014}.
Inserting the mode expansions of \refEq{eq:PBC_phi_full} and \refEq{eq:PBC_Pi2_deriv} at fine tuning, we obtain the following expression compatible with the neutrality constraints posed by the zero modes \cite{sénéchal1999introductionbosonization,Cazalilla_2004}, \begin{align}
&\langle \hat{n}(x,0)\hat{n}(y,t)\rangle \approx\nonumber
   \\
   &\Big(\frac{N}{L}\Big)^2-\frac{1}{2\pi L }\sum_{\sigma=1,2}\left[\frac{(C_{1,\sigma}+C_{2,\sigma})^2
   }{d(z_{\sigma,-},L)^2}+\frac{(C_{1,\sigma}-C_{2,\sigma})^2
   }{d(z_{\sigma,+},L)^2}\right]\nonumber
   \\
   &+\frac{1}{2}\sum_{j=1}^6\Bigg[e^{\left[\Gamma_{j,x},\Gamma_{j,y}(t)\right]/2}\cos\!\big((x-y)\mathcal A_j+t\mathcal B_j+\mathcal C_j\big)\nonumber
\\
  & \prod_{\sigma=1,2}\Big(\chi\Big)^{\eta^{\sigma}_{-,j}+\eta^{\sigma}_{+,j}}d\big(z_{\sigma,-},L\big)^{-\eta^{\sigma}_{-,j} }d\big(z_{\sigma,+},L\big)^{-\eta^{\sigma}_{+,j} }
 \Bigg]\label{correlatorperiodic},
\end{align}
with
\begin{align}
&d(x,L)=\frac{L}{\pi}\big\vert \sin\left(\frac{\pi x}{L}\right)\big\vert,\phantom{a}z_{\sigma,\pm}=x-y\pm\tilde{v}_{\sigma}t,
\end{align}
and $\chi$ is a short-distance regulator \cite{Giamarchi2003}.
The correlation function in \refEq{correlatorperiodic} embodies the spatio-temporal decay of fluctuations around the background, which depends only on light-cone coordinates $z_{\sigma,\pm}$. The chord functions $d(x,L)$ replace the algebraic decay of an infinite system, with different exponents $\eta^{\sigma}_{\pm,j}$ listed in \refTab{tab:eta_mode1} and \refTab{tab:eta_mode2} in the Appendix. The coefficients with index $j=1,2$ are of leading order with zero conformal spin, i.e., $\eta^{\sigma}_{+,j}=\eta^{\sigma}_{-,j}$, corresponding to \refEq{lowenergylongitudinaloperator}.
In this case, decay exponents are proportional to the squared diagonals of the mode decomposition matrix  $C^{(n)}_{j,\sigma}$ defined in \refEq{matrix}, where the eigenmodes are interrelated by the spatio-temporal symmetries in \refEq{parityonspinoperators} and \refEq{complexconjugationonspinoperators} respectively. More precisely, the coefficients for $j=1,2$ are permuted between the eigenmodes $\sigma=1,2$.
The indices $j\geq 3$ on the other hand, correspond to the subleading corrections in \refEq{lowenergylongitudinaloperator}, with non-zero conformal spin due to a mixing of $\hat{\phi}_j$ and  $\hat{\theta}_j$ fields in the correlation function. Consequently, these terms lead to a asymmetrical exponents in the energy and momentum resolved structure factor, while parity and conjugation symmetries according to \refEq{parityonspinoperators} and \refEq{complexconjugationonspinoperators} remain intact. For these higher order terms, the coefficients for $j=3$ and $j=4$ are the same for both eigenmodes, whereas $j=5,6$ are permuted between $\sigma=1,2$. In addition to the algebraic decay, there are zero-mode related phase factors listed in  \refTab{tablezeromodepbc} in the Appendix, that lead to energy ($t$) and momentum ($x-y$) shifts in Fourier transformed observables. Furthermore, there are phase factors that are associated to linear combinations of field commutators $\left[\Gamma_{j,x},\Gamma_{j,y}(t)\right]$, see \refTab{tablezeromodepbccommutator} in the Appendix for the complete list. This type of phase embodies branch cut discontinuities of the fundamental field correlators and ensures correct signs under particle exchange. \\
\\
In contrast to the periodic case, the universal parameters of the Dirichlet problem are actually restricted by the fine-tuning condition in \refEq{Dirichletboundaryissue3}, while the correlation functions are generally more complicated, depending on both spatial points $x$ and $y$ instead of only their difference. Using \refEq{Dirichletphi} and \refEq{Dirichletpi} we deduce the density-density correlation function from \refEq{batistaderivation}, in agreement with the neutrality condition posed by the zero modes. In contrast to the periodic case in \refEq{correlatorperiodic}, we consider only the fluctuations of the density in order to eliminate disconnected pieces, i.e.,
\begin{align}
&\langle \hat{n}(x,0)\hat{n}(y,t)\rangle-\langle\hat{n}(x,0)\rangle\langle\hat{n}(y,t)\rangle =\nonumber
   \\
   &\approx\sum_{i,j=1}^2\frac{\pi^2}{L^2}\mathrm{Cov}(\hat{N}_i,\hat{N}_j)\nonumber
   \\
   &-\sum_{\sigma=1}^2\sum_{\epsilon_1,\epsilon_2=\pm}\frac{v_{\sigma}K_{\sigma}}{4\pi^2 \bar{v}_\sigma}\sum_{\tilde{x}_{\sigma}\in\big[x+\epsilon_1y+\epsilon_2\bar{v}_{\sigma}t\big]}\frac{1}{d(\tilde{x}_{\sigma},2L)^2}\nonumber
   \\
   &+\frac{1}{2\pi^2} \sum_{j=1}^2 \sum_{\tilde{\sigma}=\pm 1} e^{-\tilde{\sigma}\left[\Gamma_{j,x},\Gamma_{j,y}(t)\right]/2} \bigg\langle \cos\bigg(\Gamma_{j,x}+\tilde{\sigma}\Gamma_{j,y}(t)\bigg)\bigg\rangle \nonumber
   \\
   &+ \frac{1}{2\pi^2} \sum_{j=3}^6 e^{\left[\Gamma_{j,x},\Gamma_{j,y}(t)\right]/2}\bigg\langle \cos\bigg(\Gamma_{j,x}-\Gamma_{j,y}(t)\bigg)\bigg\rangle.\label{correlatorobc}
\end{align}
where we have used the following vectorial notation,
\begin{align}
&\Gamma_{j,z}=\vec{a}_j\cdot \vec{X}(z,t)+\tilde{Q}_jz\label{definitionsgamma0}
\\
&\vec{X}(z,t)
\equiv
\big(\hat{\phi}_1(z,t),\ \hat{\theta}_1(z,t),\ \hat{\phi}_2(z,t),\ \hat{\theta}_2(z,t)\big)^{T},
 \nonumber \\
    &\tilde{Q} = \left(0, 0,  2Q, -2Q,  2Q, -2Q\right),\nonumber
    \\
    & \vec a^T_1=(2,0,0,0), \vec a^T_2=(0,0,2,0), \vec a^T_3=(1,-1,1,1),\nonumber
   \\
   &\vec a^T_4=(1,1,1,-1), \vec a^T_5=(1,-1,-1,1), \vec a^T_6=(1,1,-1,-1)\nonumber.
\end{align}
The overall partition of \refEq{correlatorobc} is basically analogous to the periodic case in \refEq{correlatorperiodic}, containing a zero-mode contribution followed by long-wavelength fluctations and finally by higher harmonics with decay exponents depending on the universal parameters of the theory. However, correlation functions are not exclusively dependent on light-cone coordinates, but show more complicated behavior.
This is true for the long-wavelength part, with decay exponent equal to the space-time dimension, and the higher harmonics as well.
The lowest order contributions of the higher harmonics correspond to \refEq{lowenergylongitudinaloperator} and are given by the following expression with indices $j=1,2$,
\begin{align}
           & \label{lowestorderobccorrelator}\bigg\langle \cos\bigg(\Gamma_{j,x}+\tilde{\sigma}\Gamma_{j,y}(t)\bigg)\bigg\rangle
            \propto A^{(j)}_{0,\tilde{\sigma}}\prod_{\sigma=1,2}A^{(j)}_{\sigma} B^{(j)}_{\sigma,\tilde{\sigma}}
            \end{align}
           where
            \begin{align}
            &A^{(j)}_{0,\tilde{\sigma}}=\bigg\langle \cos\bigg(\Gamma_{j,x}+\tilde{\sigma}\Gamma_{j,y}(t)\bigg)\bigg\rangle_0 \, \nonumber,
            \\
            &A^{(j)}_{\sigma}=\left[d(x,L)d(y,L)\right]^{\frac14\Delta^{(\sigma)}_j}\nonumber \,,
             \\
            &B^{(j)}_{\sigma,\tilde{\sigma}}=
            \Bigg[
            \frac{
            d(x+y-\bar{v}_\sigma t,2L)\,d(x+y+\bar{v}_\sigma t,2L)
            }{
            d(x-y-\bar{v}_\sigma t,2L)\,d(x-y+\bar{v}_\sigma t,2L)
            }\nonumber
            \Bigg]^{\frac14\,\tilde{\sigma}\Delta^{(\sigma)}_j}.
        \end{align}
Also in \refEq{lowestorderobccorrelator} we can distinguish phase factors related to zero modes $A_0$, i.e., defined in the Appendix according to \refTab{tableobczeromodes1}, from the characteristic functional decay with coefficients shown in \refTab{tab:Deltas_obc_combined}. In analogy to the periodic case, the $j=1,2$ exponents of the eigenmode channels $\sigma=1,2$ are related by the spatio-temporal symmetries in \refEq{parityonspinoperators} and \refEq{complexconjugationonspinoperators}. Phase factors originating from the commutation of the vector operators $\Gamma_{j,z}$, are defined in \refSec{subsec: Commutator Phasesobc} in the Appendix. In analogy to the periodic case, the temporal dependence in phase factors corresponds to finite-size energy shifts in Fourier resolved observables. However, due to the $\tilde{\sigma}=-1$ contribution, the spatial dependence is not only in terms of a relative coordinate $x-y$, and therefore cannot be simply associated with momentum shifts.
\\
\\
Additionally, there are also higher order contributions, corresponding to subleading corrections to \refEq{lowenergylongitudinaloperator}, that mix the dual fields with indices $3 \leq i,j\leq 6$,
\begin{align}
\label{higherorderobc}
       & \bigg\langle \cos\bigg(\Gamma_{j,x}-\Gamma_{j,y}(t)\bigg)\bigg\rangle\propto \tilde{A}^{(j)}_0\prod_{\sigma=1}^2\tilde{A}^{(j)}_{\sigma}\tilde{B}^{+,j}_{\sigma}\tilde{B}^{-,j}_{\sigma}\tilde{C}^{(j)}_{\sigma},
\end{align}
with
\begin{align}
&\tilde{A}^{(j)}_0=\bigg\langle \cos\bigg(\Gamma_{j,x}-\Gamma_{j,y}(t)\bigg)\bigg\rangle_0,
\,  
\\
&\tilde{A}^{(j)}_{\sigma}=[d(x,L)d(y,L)]^{\frac14 (\Delta_{b,j}^{(\sigma)}-\Delta_{a,j}^{(\sigma)})},
\nonumber
\\
&\tilde{B}^{+,j}_{\sigma}=
\Big[d(x+y-\bar{v}_\sigma t,2L)d(x+y+\bar{v}_\sigma t,2L)\Big]^{
\frac14(\Delta_{a,j}^{(\sigma)}-\Delta_{b,j}^{(\sigma)})},
\nonumber
\\
&\tilde{B}^{-,j}_{\sigma}=
\Big[d(x-y-\bar{v}_\sigma t,2L)d(x-y+\bar{v}_\sigma t,2L)\Big]^{-
\frac14(\Delta_{a,j}^{(\sigma)}+\Delta_{b,j}^{(\sigma)})},
\nonumber
\\
&
\tilde{C}^{(j)}_{\sigma}=
\bigg[\frac{d(x-y+\bar{v}_\sigma t,2L)}{d(x-y-\bar{v}_\sigma t,2L)}\bigg]^{\frac12\sqrt{\Delta_{a,j}^{(\sigma)}\Delta_{b,j}^{(\sigma)}}}.\nonumber,
\end{align}
where decay coefficients are given in  \refTab{tab:Deltas_obc_combined}  and zero mode factors in \refTab{tableobczeromodes2} in the Appendix. In analogy to the periodic case, the coefficients are related by spatio-temporal symmetry under the premise of \refEq{Dirichletboundaryissue3}. Furthermore, the chiral imbalance also leads to asymmetric exponents in the corresponding spectroscopic probes, whereas the full signal is spatio-temporally symmetric. 
\\
\\
Ultimately, the fine tuning condition in \refEq{Dirichletboundaryissue3} allows for a decoupling of the chains into two independent Luttinger liquids. Consequently, we expect that correlation functions in the bulk, i.e., $\big\vert (x-y-\bar{v}_{\sigma t})(x-y+\bar{v}_{\sigma t})\big\vert\ll xy$, behave differently compared to those near boundaries, $\vert\bar{v}_{\sigma}t\vert\gg x,y$ \cite{Cardy1984,CARDY1989581,Recknagel_Schomerus_2013}. However, it should be stressed that the bulk limit can only coincide with that of the cyclic case under fine-tuning, which is not a necessity for periodic boundary conditions. In order to test bulk and boundary behavior explicitly for our model, we calculate the bulk and boundary limits for \refEq{correlatorobc}. The long wavelength fluctuations already reveal the emergence of translational invariance in the bulk, expressed suitably in terms of light-cone coordinates $z_{\sigma,\tilde{\sigma}}$, with a scaling dimension equal to the space time dimension,
\begin{align} 
    &\sum_{\tilde{x}_{\sigma}\in\big[x+\epsilon_1y+\epsilon_2\tilde{v}_{\sigma}t\big]}\frac{1}{d(\tilde{x}_{\sigma},2L)^2},\phantom{a}z_{\sigma,\tilde{\sigma}}=x-y+\tilde{\sigma}\tilde{v}_{\sigma}t\nonumber
    \\
    &\propto\begin{cases}
\sum_{\sigma=1,2}\sum_{\tilde{\sigma}=\pm 1}\vert z_{\sigma,\tilde{\sigma}}\vert^{-2}, \phantom{a}\vert z_{\sigma,+}z_{\sigma,-}\vert\ll xy\\
            \sum_{\sigma=1,2} \vert \tilde{v}_{\sigma} t\vert^{-2}, \phantom{a}\vert \tilde{v}_{\sigma}t\vert\gg x,y,\phantom{a}x\approx y\end{cases}
\end{align}
The bulk decay is proportional to the long wavelength fluctuation of the  periodic case in \refEq{correlatorperiodic} at fine-tuning, whereas the boundary limit does not have an analog in the cyclic system.
For the lowest order vertex operators in \refEq{lowestorderobccorrelator} with $ j=1,2$, the $\tilde{\sigma}=1$ part vanishes asymptotically, whereas the $\tilde{\sigma}=-1$ contribution corresponds to the infinite size limit of \refEq{correlatorperiodic} at fine-tuning,
\begin{align}
             & \bigg\langle \cos\bigg(\Gamma_{j,x}+\tilde{\sigma}\Gamma_{j,y}\bigg)\bigg\rangle\overbrace{\propto}^{\vert z_{\sigma,+}z_{\sigma,-}\vert\ll xy}\nonumber
             \\
             &\begin{cases}
        0, \phantom{a}\tilde{\sigma}=+1 \\
          \vert z_{j,+} z_{j,-}\vert^{-\frac{v_{j}K_{j}}{\bar{v}_{j}}}, \phantom{a}\tilde{\sigma}=-1,
         \end{cases}\label{bulklimitlowestorder}
        \end{align}
On the other hand, left and right movers become functionally dependent near a boundary for both choices of $\tilde{\sigma}$, and the decay coefficient is doubled compared to the bulk  \cite{Cardy1984,cardy1991a,Eggert92}
\begin{align}
     \bigg\langle \cos\bigg(\Gamma_{j,x}+\tilde{\sigma}\Gamma_{j,y}\bigg)\bigg\rangle\overbrace{\approx}^{\vert \tilde{v}_{\sigma}t\vert\gg x,y,\phantom{a}x\approx y}
           \vert x y\vert^{-\frac{v_{j}K_{j}}{\bar{v}_{j}}}.
\end{align}
Similar behaviour is found in the asymptotics of the sub-leading terms in \refEq{higherorderobc}, i.e.,
\begin{align}
 & \bigg\langle \cos\bigg(\Gamma_{j,x}-\Gamma_{j,y}\bigg)\bigg\rangle \nonumber
  \\
  &\propto\begin{cases}
        \prod_{\sigma=1,2}\vert z_{\sigma,+}\vert^{-\tilde{\eta}^{(\sigma)}_{+,j}}
\vert z_{\sigma,-}\vert^{-\tilde{\eta}^{(\sigma)}_{-,j}}, \phantom{a}\vert z_{\sigma,+}z_{\sigma,-}\vert\ll xy,\\
           \prod_{\sigma=1}^2\vert xy\vert^{\frac14 (\Delta_{b,j}^{(\sigma)}-\Delta_{a,j}^{(\sigma)})}\vert\tilde{v}_{\sigma}t \vert^{-\Delta_{b,j}^{(\sigma)}}, \phantom{a}\vert \tilde{v}_{\sigma}t\vert\gg x,y,\end{cases}
\end{align}
with
\begin{align}
           &\tilde{\eta}^{(\sigma)}_{+,j}=
\frac{\Delta_{a,j}^{(\sigma)}+\Delta_{b,j}^{(\sigma)}-2\sqrt{\Delta_{a,j}^{(\sigma)}\Delta_{b,j}^{(\sigma)}}}{4} ,
\\
&\tilde{\eta}^{(\sigma)}_{-,j}=
\frac{\Delta_{a,j}^{(\sigma)}+\Delta_{b,j}^{(\sigma)}+2\sqrt{\Delta_{a,j}^{(\sigma)}\Delta_{b,j}^{(\sigma)}}} {4}. \nonumber
\end{align}
It should be noted that left and right movers decay with different exponents in the bulk, due to the mixing of density and phase fields in the correlation function. Again, the bulk limit coincides with that of the sub-leading contributions of \refEq{correlatorperiodic}, under the premise of \refEq{Dirichletboundaryissue3}.

\begin{table}[t]
\centering
\renewcommand{\arraystretch}{1.25}
\setlength{\tabcolsep}{10pt}

\begin{tabular}{c|cc}
\multicolumn{3}{c}{\textbf{Leading order}} \\
\hline
$j$ &
$\displaystyle \Delta^{(1)}_j$ &
$\displaystyle \Delta^{(2)}_j$ \\
\hline
1 &
$\displaystyle -\frac{4v_1 K_1}{\bar{v}_1}$ &
$0$ \\[2mm]
2 &
$0$ &
$\displaystyle -\frac{4v_2 K_2}{\bar{v}_2}$ \\
\end{tabular}

\vspace{4mm}

\begin{tabular}{c|cc}
\multicolumn{3}{c}{\textbf{Subleading order \(\tilde v_1\)}} \\
\hline
$j$ &
$\displaystyle \Delta_{a,j}^{(1)}$ &
$\displaystyle \Delta_{b,j}^{(1)}$ \\
\hline
3 &
$\displaystyle \frac{v_1 K_1}{\tilde{v}_1}
\Big(1-\frac{\alpha_2}{\pi}\Big)^2$ &
$\displaystyle \frac{\tilde{v}_1}{v_1 K_1}$ \\[2mm]
4 &
$\displaystyle \frac{v_1 K_1}{\tilde{v}_1}
\Big(1+\frac{\alpha_2}{\pi}\Big)^2$ &
$\displaystyle \frac{\tilde{v}_1}{v_1 K_1}$ \\[2mm]
5 &
$\displaystyle \frac{v_1 K_1}{\tilde{v}_1}
\Big(1-\frac{\alpha_2}{\pi}\Big)^2$ &
$\displaystyle \frac{\tilde{v}_1}{v_1 K_1}$ \\[2mm]
6 &
$\displaystyle \frac{v_1 K_1}{\tilde{v}_1}
\Big(1+\frac{\alpha_2}{\pi}\Big)^2$ &
$\displaystyle \frac{\tilde{v}_1}{v_1 K_1}$ \\
\end{tabular}

\vspace{4mm}

\begin{tabular}{c|cc}
\multicolumn{3}{c}{\textbf{Subleading order \(\tilde v_2\)}} \\
\hline
$j$ &
$\displaystyle \Delta_{a,j}^{(2)}$ &
$\displaystyle \Delta_{b,j}^{(2)}$ \\
\hline
3 &
$\displaystyle \frac{v_2 K_2}{\tilde{v}_2}
\Big(1+\frac{\alpha_1}{\pi}\Big)^2$ &
$\displaystyle \frac{\tilde{v}_2}{v_2 K_2}$ \\[2mm]
4 &
$\displaystyle \frac{v_2 K_2}{\tilde{v}_2}
\Big(1-\frac{\alpha_1}{\pi}\Big)^2$ &
$\displaystyle \frac{\tilde{v}_2}{v_2 K_2}$ \\[2mm]
5 &
$\displaystyle \frac{v_2 K_2}{\tilde{v}_2}
\Big(1-\frac{\alpha_1}{\pi}\Big)^2$ &
$\displaystyle \frac{\tilde{v}_2}{v_2 K_2}$ \\[2mm]
6 &
$\displaystyle \frac{v_2 K_2}{\tilde{v}_2}
\Big(1+\frac{\alpha_1}{\pi}\Big)^2$ &
$\displaystyle \frac{\tilde{v}_2}{v_2 K_2}$ \\
\end{tabular}

\caption{Decay exponents for the leading order contributions of \refEq{lowestorderobccorrelator} and the subleading contributions in \refEq{higherorderobc}, for the open boundary density-density correlator in \refEq{correlatorobc}. 
$\alpha_{1,2}$ are the flux-tube constants defined in \refEq{fluxtubeattachement}
For \(j=3,4,5,6\), the entries correspond to \(i=j\). Due to the mixing of $\hat{\phi}_j$ and $\hat{\theta}_j$ fields, the higher order correlators have a non-zero conformal spin.}
\label{tab:Deltas_obc_combined}
\end{table}

\section{Conclusion and Outlook}
\label{sec: conclusion}

We provide the diagonalization of
a two mode Schulz-Shastry model for periodic and Dirichlet boundary conditions, which realizes anyonic excitations on top of the helical ground states of a spin 1/2 zigzag chain. We find that the particle current across a boundary is only vanishing whenever the coupling constants  of the model can be rephrased in terms of a free anyonic theory. This is reminiscent of the first introduction of 1d anyons of Leinaas and Myrheim \cite{Leinaas1977}, where the statistical angle parametrizes boundary conditions at coincidence points in configuration space. Statistical interactions inextricably couple the finite-size spectrum of the underlying conformal field theories, which leads to commensurability effects in persistent currents that are intertwined with the parity of the total number zero mode. At the level of excitations, statistical transmutation acts by reducing the velocities of density excitations by mutual interactions, where the free boson limit is attained whenever they vanish completely. This transmutation can be probed transparently with statistically invariant variables. To this end, we have investigated  Friedel oscillations for open boundary conditions that distinguish between bosonic and fermionic exclusion statistics. Additionally, we find that these oscillations are modulated differently for an even or odd number of excitations above the ground state. In contrast to other variants of anyonic Luttinger liquids, exchange statistic is modified in between the two chains, not inside each chain individually. Likewise, there is no intrinsic spatio-temporal asymmetry for inter-chain anyons and chiral species have equal velocities for each eigenmode. Interestingly, we find chirally imbalanced, subleading contributions for each eigenmode in longitudinal spin correlations, that are paired with the other chain in order to restore spatio-temporal symmetry globally. These asymmetric exponents could be revealed by the corresponding spectroscopic measures. \\
\\
Beyond the properties of free field theories studied here, it would be interesting to include interaction effects and study the impact of statistical transmutations on phase transitions. Furthermore, the investigation of non-linear effects for this kind of Luttinger liquid remains a formidable task.

\section{Acknowledgement}
\label{sec: Acknowledgement}
The authors thank Volker Schomerus and Christian D. Batista for invaluable discussions.     
M.B.\ and T.P.\ acknowledge
funding by the European Union (ERC, QUANTWIST, project number $101039098$). The views and opinions
expressed are however those of authors only and do
not necessarily reflect those of the European Union
or the European Research Council, Executive Agency.
T.P.\ acknowledges support by the Cluster of Excellence “CUI: Advanced Imaging of Matter” of the Deutsche Forschungsgemeinschaft (DFG) – EXC
2056 – project ID $390715994$. 
B.P. is supported by the Partnership for Innovation,
Education and Research (PIER) between Deutsches
Elektronen-Synchrotron (DESY) and Universität Hamburg (UHH), as well as the MIT Global Experience
(MISTI) program.

\appendix
\setcounter{equation}{0}   
\renewcommand{\theequation}{S\arabic{equation}}  
\setcounter{table}{0}   
\renewcommand{\thetable}{S\arabic{table}}  
\setcounter{section}{0}   
\renewcommand{\thesection}{S\arabic{section}}  
\clearpage
\onecolumngrid

\clearpage
\onecolumngrid

\section*{Supplemental Material: }
\label{sec: supplemental}

In this supplementary, we define the basic constituents of the density-density correlation function in \refEq{correlatorperiodic} and \refEq{correlatorobc}, which are only abbreviated in the main text, for the sake of readability. The calculation of the correlator per se follows standard methods \cite{Eggert2003,Giamarchi2003,sénéchal1999introductionbosonization}, where we consider the low-energy expansion around two different Fermi-points in \refEq{lowenergyürojection} for the fermionic density operator $\hat{n}_j=\hat{c}^{\dagger}_j\hat{c}_j$ and subsequently insert the boundary-specific mode expansions in  \refEq{eq:PBC_phi_full} and \refEq{eq:PBC_Pi2_deriv} for the cyclic system, or \refEq{Dirichletphi} and \refEq{Dirichletpi} for the Dirichlet case. In order to compare the boundary conditions explicitly, and physically interpret the correlation function for the periodic case, we consider demand a fine tuning of constants according to  \refEq{Dirichletboundaryissue3} for both types. Here, only contributions to the expectation value are nonvanishing that are in accordance with the neutrality conditions of the underlying conformal field theory \cite{Cazalilla_2004}.

\subsection{Density-Density Correlation Function for Periodic Boundary Conditions}
\label{subsec: densitydensityperiodicpbc}

We list all zero-mode phases of \refEq{correlatorperiodic} in table \refTab{tablezeromodepbc}, where they share the same functional form $\cos\!\big((x-y)\mathcal A_j+t\mathcal B_j+\mathcal C_j\big)$ for all indices $j=1,2,..,6$. We see that the different phase contributions for the indices $j=1,2$ contain only a single quantum number each, where the expectation value is taken with respect to \refEq{eq: zeromodehamiltonianpbc}. The indices $j\geq3$ on the other hand involve the quantum numbers of both chains, reflecting that the contributions of different eigenmodes are mixed.
This mixing is a signature of the fact that $j\geq3$ terms correspond to higher order corrections involving density and phase fields of both sectors.

\begin{table}[!htbp]
\centering
\renewcommand{\arraystretch}{1.35}
\setlength{\tabcolsep}{4pt}
\begin{tabular*}{\linewidth}{@{\extracolsep{\fill}}c|c|c|c@{}}
$j$ & $\mathcal A_j$ & $\mathcal B_j$ & $\mathcal C_j$ \\
\hline
1 &
$\displaystyle -\frac{2\pi}{L}\langle N_1\rangle$ &
$\displaystyle \frac{g_2}{2L}\langle N_2\rangle$ &
$\displaystyle -\frac{2v_1 K_1\pi}{L}\langle J_1\rangle$ \\
\hline
2 &
$\displaystyle -\frac{2\pi}{L}\langle N_2\rangle$ &
$\displaystyle \frac{g_1}{2L}\langle N_1\rangle$ &
$\displaystyle -\frac{2v_2 K_2\pi}{L}\langle J_2\rangle$ \\
\hline
3 &
$\displaystyle 2Q+\frac{\pi}{L}\!\left(-N-\Delta_- J\right)$ &
$\displaystyle \left(-\frac{v_1\pi}{K_1 L}+\frac{g_1}{4L}\right)\langle N_1\rangle
+\left(\frac{v_2\pi}{K_2 L}+\frac{g_2}{4L}\right)\langle N_2\rangle
-\frac{g_2}{4L}\langle J_1\rangle
+\frac{g_1}{4L}\langle J_2\rangle$ &
$\displaystyle -\frac{v_1 K_1\pi}{L}\!\langle J_1\rangle
-\frac{v_2 K_2\pi}{L}\!\langle J_2\rangle$ \\
\hline
4 &
$\displaystyle -2Q+\frac{\pi}{L}\!\left(-N+\Delta_- J\right)$ &
$\displaystyle \left(\frac{v_1\pi}{K_1 L}+\frac{g_1}{4L}\right)\langle N_1\rangle
+\left(-\frac{v_2\pi}{K_2 L}+\frac{g_2}{4L}\right)\langle N_2\rangle
+\frac{g_2}{4L}\langle J_1\rangle
-\frac{g_1}{4L}\langle J_2\rangle$ &
$\displaystyle -\frac{v_1 K_1\pi}{L}\!\langle J_1\rangle
-\frac{v_2 K_2\pi}{L}\!\langle J_2\rangle$ \\
\hline
5 &
$\displaystyle 2Q+\frac{\pi}{L}\!\left(-\Delta_- N-\Delta_- J\right)$ &
$\displaystyle \left(-\frac{v_1\pi}{K_1 L}-\frac{g_1}{4L}\right)\langle N_1\rangle
+\left(\frac{v_2\pi}{K_2 L}+\frac{g_2}{4L}\right)\langle N_2\rangle
-\frac{g_2}{4L}\langle J_1\rangle
+\frac{g_1}{4L}\langle J_2\rangle$ &
$\displaystyle -\frac{v_1 K_1\pi}{L}\!\langle J_1\rangle
+\frac{v_2 K_2\pi}{L}\!\langle J_2\rangle$ \\
\hline
6 &
$\displaystyle -2Q+\frac{\pi}{L}\!\left(-\Delta_- N+\Delta_- J\right)$ &
$\displaystyle \left(\frac{v_1\pi}{K_1 L}-\frac{g_1}{4L}\right)\langle N_1\rangle
+\left(-\frac{v_2\pi}{K_2 L}+\frac{g_2}{4L}\right)\langle N_2\rangle
+\frac{g_2}{4L}\langle J_1\rangle
-\frac{g_1}{4L}\langle J_2\rangle$ &
$\displaystyle -\frac{v_1 K_1\pi}{L}\!\langle J_1\rangle
+\frac{v_2 K_2\pi}{L}\!\langle J_2\rangle$ \\
\hline
\end{tabular*}
\caption{Zero-mode contribution $\langle\cos(\Gamma_{i,x}-\Gamma_{i,y})_{\mathrm{zero}}\rangle
=\cos\!\big((x-y)\mathcal A_i+t\mathcal B_i+\mathcal C_i\big)$ for $i=1,\dots,6$; of the density-density correlator im \refEq{correlatorperiodic}. The phase can be divided into spatial, temporal and constant contributions, corresponding to shifts in momentum and energy, or just a global phase. It should be noted that we have defined the short hand notation $\Delta_\pm X=\langle X_1\rangle\pm\langle X_2\rangle$ for the quantum numbers, where $N=\langle \hat{N}_1+\hat{N}_2\rangle$ denotes the total number of excitations.}
\label{tablezeromodepbc}
\label{tab:PBC_cos_zero_decomp_xy}
\end{table}

Analogously, the intertwining of different fields is represented as well by the scaling dimensions of the two eigenmodes in \refTab{tab:eta_mode1} and \refTab{tab:eta_mode2}. Here, only the $j=1,2$ contributions have a vanishing conformal spin, while it is nonzero for the other indices $j=3,4,5,6$.

\begin{table}[!htbp]
\centering
\scriptsize
\renewcommand{\arraystretch}{1.25}
\setlength{\tabcolsep}{8pt}
\begin{tabular*}{\linewidth}{@{\extracolsep{\fill}}c|c|c|c@{}}
$j$ & $\eta^{(1)}_{-,j}$ & $\eta^{(1)}_{+,j}$ & $s_{1,j}$ \\
\hline
1 &
$\displaystyle \frac{  v_1K_1}{ \tilde v_1}$ &
$\displaystyle \frac{  v_1K_1}{ \tilde v_1}$ &
$0$ \\[2mm]
2 &
$0$ &
$0$ &
$0$ \\[2mm]
\hline
3 &
$\displaystyle
\frac{ v_1K_1}{ 4\tilde v_1}
\left(1+\frac{\alpha_2}{\pi}+\frac{\tilde v_1}{v_1 K_1}\right)^2$ &
$\displaystyle
\frac{ v_1K_1}{ 4\tilde v_1}
\left(1+\frac{\alpha_2}{\pi}-\frac{\tilde v_1}{v_1 K_1}\right)^2$ &
$\displaystyle
\frac{  v_1K_1}{ 2\tilde v_1}
\left(1+\frac{\alpha_2}{\pi}\right)\frac{\tilde v_1}{v_1 K_1}$ \\[2mm]
4 &
$\displaystyle
\frac{ v_1K_1}{ 4\tilde v_1}
\left(1-\frac{\alpha_2}{\pi}-\frac{\tilde v_1}{v_1 K_1}\right)^2$ &
$\displaystyle
\frac{ v_1K_1}{ 4\tilde v_1}
\left(1-\frac{\alpha_2}{\pi}+\frac{\tilde v_1}{v_1 K_1}\right)^2$ &
$\displaystyle
-\frac{  v_1K_1}{ 2\tilde v_1}
\left(1-\frac{\alpha_2}{\pi}\right)\frac{\tilde v_1}{v_1 K_1}$ \\[2mm]
5 &
$\displaystyle
\frac{ v_1K_1}{ 4\tilde v_1}
\left(1+\frac{\alpha_2}{\pi}+\frac{\tilde v_1}{v_1 K_1}\right)^2$ &
$\displaystyle
\frac{ v_1K_1}{ 4\tilde v_1}
\left(1+\frac{\alpha_2}{\pi}-\frac{\tilde v_1}{v_1 K_1}\right)^2$ &
$\displaystyle
\frac{  v_1K_1}{ 2\tilde v_1}
\left(1+\frac{\alpha_2}{\pi}\right)\frac{\tilde v_1}{v_1 K_1}$ \\[2mm]
6 &
$\displaystyle
\frac{ v_1K_1}{ 4\tilde v_1}
\left(1-\frac{\alpha_2}{\pi}-\frac{\tilde v_1}{v_1 K_1}\right)^2$ &
$\displaystyle
\frac{ v_1K_1}{ 4\tilde v_1}
\left(1-\frac{\alpha_2}{\pi}+\frac{\tilde v_1}{v_1 K_1}\right)^2$ &
$\displaystyle
-\frac{  v_1K_1}{2 \tilde v_1}
\left(1-\frac{\alpha_2}{\pi}\right)\frac{\tilde v_1}{v_1 K_1}$ \\
\hline
\end{tabular*}
\caption{Decay coefficients of mode $\tilde v_1$ in \refEq{correlatorperiodic} and conformal spin $s_{1,j}=(\eta^{(1)}_{-,j}-\eta^{(1)}_{+,j})/2$, where universal constants obey the fine tuning in \refEq{Dirichletboundaryissue3}. Here, $\alpha_{1,2}$ are the flux-tube constants defined in \refEq{fluxtubeattachement}. The indices $j=1,2$ correspond to the lowest order contributions, while $j\geq 3$ embody subleading corrections.}
\label{tab:eta_mode1}
\end{table}

\begin{table}[!htbp]
\centering
\scriptsize
\renewcommand{\arraystretch}{1.25}
\setlength{\tabcolsep}{8pt}
\begin{tabular*}{\linewidth}{@{\extracolsep{\fill}}c|c|c|c@{}}
$j$ & $\eta^{(2)}_{-,j}$ & $\eta^{(2)}_{+,j}$ & $s_{2,j}$ \\
\hline
1 &
$0$ &
$0$ &
$0$ \\[2mm]
2 &
$\displaystyle \frac{ v_2K_2}{ \tilde v_2}$ &
$\displaystyle \frac{ v_2K_2}{ \tilde v_2}$ &
$0$ \\[2mm]
\hline
3 &
$\displaystyle
\frac{v_2K_2}{4 \tilde v_2}
\left(1+\frac{\alpha_2}{\pi}+\frac{\tilde v_2}{v_2 K_2}\right)^2$ &
$\displaystyle
\frac{v_2K_2}{4 \tilde v_2}
\left(1+\frac{\alpha_2}{\pi}-\frac{\tilde v_2}{v_2 K_2}\right)^2$ &
$\displaystyle
\frac{v_2K_2}{ 2\tilde v_2}
\left(1+\frac{\alpha_2}{\pi}\right)\frac{\tilde v_2}{v_2 K_2}$ \\[2mm]
4 &
$\displaystyle
\frac{v_2K_2}{4 \tilde v_2}
\left(1-\frac{\alpha_2}{\pi}-\frac{\tilde v_2}{v_2 K_2}\right)^2$ &
$\displaystyle
\frac{v_2K_2}{4 \tilde v_2}
\left(1-\frac{\alpha_2}{\pi}+\frac{\tilde v_2}{v_2 K_2}\right)^2$ &
$\displaystyle
-\frac{v_2K_2}{2 \tilde v_2}
\left(1-\frac{\alpha_2}{\pi}\right)\frac{\tilde v_2}{v_2 K_2}$ \\[2mm]
5 &
$\displaystyle
\frac{v_2K_2}{4 \tilde v_2}
\left(1-\frac{\alpha_2}{\pi}-\frac{\tilde v_2}{v_2 K_2}\right)^2$ &
$\displaystyle
\frac{v_2K_2}{4 \tilde v_2}
\left(1-\frac{\alpha_2}{\pi}+\frac{\tilde v_2}{v_2 K_2}\right)^2$ &
$\displaystyle
-\frac{v_2K_2}{2 \tilde v_2}
\left(1-\frac{\alpha_2}{\pi}\right)\frac{\tilde v_2}{v_2 K_2}$ \\[2mm]
6 &
$\displaystyle
\frac{v_2K_2}{4 \tilde v_2}
\left(1+\frac{\alpha_2}{\pi}+\frac{\tilde v_2}{v_2 K_2}\right)^2$ &
$\displaystyle
\frac{v_2K_2}{4 \tilde v_2}
\left(1+\frac{\alpha_2}{\pi}-\frac{\tilde v_2}{v_2 K_2}\right)^2$ &
$\displaystyle
\frac{v_2K_2}{ 2\tilde v_2}
\left(1+\frac{\alpha_2}{\pi}\right)\frac{\tilde v_2}{v_2 K_2}$ \\
\hline
\end{tabular*}
\caption{Decay coefficients of mode $\tilde v_2$ in \refEq{correlatorperiodic} and conformal spin $s_{2,j}=(\eta^{(2)}_{-,j}-\eta^{(2)}_{+,j})/2$ for the fine tuned condition in \ref{Dirichletboundaryissue3}. Here $\alpha_{1,2}$ are the flux-tube constants defined in \refEq{fluxtubeattachement}. In analogy to \refTab{tab:eta_mode1}, the indices $j=1,2$ correspond to the lowest order contributions, while $j\geq 3$ embody subleading corrections.}
\label{tab:eta_mode2}
\end{table}

\subsubsection{Commutator Phases}\label{subsec: Commutator Phasesperiodic}

In this section, we list the definitions for the calculation of commutator-related phases $\left[\Gamma_{j,x},\Gamma_{j,y}(t)\right]$ for the periodic density-density correlator in \refEq{correlatorperiodic}. 
To this end, we use the vectorial notation of fields, in \refEq{definitionsgamma0}, i.e., 
\begin{align}
&\Gamma_{j,z}=\vec{a}_j\cdot \vec{X}(z,t)+\tilde{Q}_jz\label{definitionsgamma0}
\\
&\vec{X}(z,t)
\equiv
\big(\hat{\phi}_1(z,t),\ \hat{\theta}_1(z,t),\ \hat{\phi}_2(z,t),\ \hat{\theta}_2(z,t)\big)^{T},
 \nonumber \\
    &\tilde{Q} = \left(0, 0,  2Q, -2Q,  2Q, -2Q\right),\nonumber
    \\
    & \vec a^T_1=(2,0,0,0), \vec a^T_2=(0,0,2,0), \vec a^T_3=(1,-1,1,1),\nonumber
   \\
   &\vec a^T_4=(1,1,1,-1), \vec a^T_5=(1,-1,-1,1), \vec a^T_6=(1,1,-1,-1)\nonumber.
\end{align}
that we can further additively decompose into the zero mode contribution $\left[\Gamma_{j,x},\Gamma_{j,y}(t)\right]_{\mathrm{zero}}$ and the oscillating part  $\left[\Gamma_{j,x},\Gamma_{j,y}(t)\right]_{\mathrm{osc}}$. 

\begin{table}[!htbp] 
\centering
\scriptsize
\renewcommand{\arraystretch}{1.25}
\setlength{\tabcolsep}{6pt}
\begin{tabular}{c|l}
$i$ & $\tfrac{1}{2}[\Gamma_i(x,0),\Gamma_i(y,t)]_{0}$ \\
\hline
1 & $\displaystyle i\,t\,\frac{2v_1\pi}{K_1L}$ \\
2 & $\displaystyle i\,t\,\frac{2v_2\pi}{K_2L}$ \\
3 & $\displaystyle i\,t\left( \frac{v_1\pi}{2K_1L} +\frac{v_1K_1\pi}{2L} +\frac{v_2\pi}{2K_2L} +\frac{v_2K_2\pi}{2L} -\frac{g_1-g_2}{4L} \right)$ \\
4 & $\displaystyle i\,t\left( \frac{v_1\pi}{2K_1L} +\frac{v_1K_1\pi}{2L} +\frac{v_2\pi}{2K_2L} +\frac{v_2K_2\pi}{2L} +\frac{g_1-g_2}{4L} \right)$ \\
5 & $\displaystyle i\,t\left( \frac{v_1\pi}{2K_1L} +\frac{v_1K_1\pi}{2L} +\frac{v_2\pi}{2K_2L} +\frac{v_2K_2\pi}{2L} +\frac{g_1+g_2}{4L} \right) \;+\; i\,\frac{2\pi}{L}(x-y)$ \\
6 & $\displaystyle i\,t\left( \frac{v_1\pi}{2K_1L} +\frac{v_1K_1\pi}{2L} +\frac{v_2\pi}{2K_2L} +\frac{v_2K_2\pi}{2L} -\frac{g_1+g_2}{4L} \right) \;-\; i\,\frac{2\pi}{L}(x-y)$ \\
\hline
\end{tabular}
\caption{Zero-mode commutators $\tfrac{1}{2}[\Gamma_i(x,0),\Gamma_i(y,t)]_{\rm zero}$ for \refEq{correlatorperiodic}. 
The components are defined via \refEq{definitionsgamma0} as well as \refEq{eq:PBC_phi_full} and \refEq{eq:PBC_Pi2_deriv}, yielding phase shifts in time and relative coordinate $x-y$.}
\label{tablezeromodepbccommutator}
\end{table}

The zero mode part is listed in \refTab{tablezeromodepbccommutator}, resulting in energy and momentum shifts as well. However, these shifts are purely constants, independent of the zero mode eigenvalues in contrast to \refTab{tablezeromodepbc}.
The calculation of the oscillating part of the commutators is more involved, where we use the following notation for the field commutators,
\begin{align}
F_{\mu\nu}(\Delta,t)
&\equiv
\big[\mathbf X_\mu(x,0),\mathbf X_\nu(y,t)\big]_{\mathrm{osc}},
\qquad
\mu,\nu\in\{1,2,3,4\}, \qquad
\Delta\equiv x-y.
\label{def:Fmunu}
\end{align}
For the oscillator part of  commutators we find the general relations,
\begin{align}
F_{\mu\nu}(-\Delta,-t)
&=
-F_{\mu\nu}(\Delta,t),\nonumber
\\
&
\\
F_{\nu\mu}(\Delta,t)\nonumber
&=
-F_{\mu\nu}(-\Delta,-t)
=
F_{\mu\nu}(\Delta,t).
\end{align}
Hence, the diagonal oscillator commutator can be written as
\begin{align}
[\Gamma_j(x,0),\Gamma_j(y,t)]_{\mathrm{osc}}
&=
(a_j^1)^2F_{11}
+(a_j^2)^2F_{22}
+(a_j^3)^2F_{33}
+(a_j^4)^2F_{44}
\nonumber\\
&\quad
+2a_j^1a_j^2F_{12}
+2a_j^1a_j^3F_{13}
+2a_j^1a_j^4F_{14}
\nonumber\\
&\quad
+2a_j^2a_j^3F_{23}
+2a_j^2a_j^4F_{24}
+2a_j^3a_j^4F_{34}.
\label{Gamma_diag_osc_general}
\end{align}
where all \(F_{\mu\nu}\) on the right-hand side are evaluated at \((\Delta,t)\).

In \refEq{Gamma_diag_osc_general} $a^{(l)}_m$ denotes the $l$-th component of the $m$-th vector in \refEq{definitionsgamma0}. The different components for $j\in[1,6]$ are then given by the following linear combinations of basic commutators in \refEq{def:Fmunu},
\begin{align}
[\Gamma_1(x,0),\Gamma_1(y,t)]_{\mathrm{osc}}
&=
4F_{11},
\label{Gamma_diag_osc_1}
\\[4pt]
[\Gamma_2(x,0),\Gamma_2(y,t)]_{\mathrm{osc}}
&=
4F_{33},
\label{Gamma_diag_osc_2}
\\[6pt]
[\Gamma_3(x,0),\Gamma_3(y,t)]_{\mathrm{osc}}
&=
F_{11}+F_{22}+F_{33}+F_{44}
-2F_{12}
+2F_{13}
+2F_{14}
-2F_{23}
-2F_{24}
+2F_{34},
\label{Gamma_diag_osc_3}
\\[6pt]
[\Gamma_4(x,0),\Gamma_4(y,t)]_{\mathrm{osc}}
&=
F_{11}+F_{22}+F_{33}+F_{44}
+2F_{12}
+2F_{13}
-2F_{14}
+2F_{23}
-2F_{24}
-2F_{34},
\label{Gamma_diag_osc_4}
\\[6pt]
[\Gamma_5(x,0),\Gamma_5(y,t)]_{\mathrm{osc}}
&=
F_{11}+F_{22}+F_{33}+F_{44}
-2F_{12}
-2F_{13}
+2F_{14}
+2F_{23}
-2F_{24}
-2F_{34},
\label{Gamma_diag_osc_5}
\\[6pt]
[\Gamma_6(x,0),\Gamma_6(y,t)]_{\mathrm{osc}}
&=
F_{11}+F_{22}+F_{33}+F_{44}
+2F_{12}
-2F_{13}
-2F_{14}
-2F_{23}
-2F_{24}
+2F_{34}.
\label{Gamma_diag_osc_6}
\end{align}

The commutators of the oscillating parts of the fields can be deduced from \refEq{eq:PBC_phi_full} and \refEq{eq:PBC_Pi2_deriv}, which yields,
\begingroup
\small
\begin{align}
\operatorname{SGN}(x,y)
&\equiv
\frac{\pi}{2}\left[
\operatorname{sgn}(x+y)+\operatorname{sgn}(x-y)
\right]
-\frac{2\pi x}{L},
\end{align}
\begin{align}
[\hat{\phi}_{\mathrm{osc},1}(x,0),\hat{\phi}_{\mathrm{osc},1}(y,t)]
&=
\frac{i}{\pi}
\frac{\pi v_1K_1}{2\tilde v_1}\,
\operatorname{SGN}(\tilde v_1 t,x-y),
\\[4pt]
[\hat{\phi}_{\mathrm{osc},1}(x,0),\hat{\phi}_{\mathrm{osc},2}(y,t)]
&=0,
\\[4pt]
[\hat{\phi}_{\mathrm{osc},2}(x,0),\hat{\phi}_{\mathrm{osc},2}(y,t)]
&=
\frac{i}{\pi}
\frac{\pi v_2K_2}{2\tilde v_2}\,
\operatorname{SGN}(\tilde v_2 t,x-y),
\\[6pt]
[\hat{\phi}_{\mathrm{osc},1}(x,0),\hat{\theta}_{\mathrm{osc},1}(y,t)]
&=
\frac{i}{\pi}
\frac{\tilde v_1\frac{\pi v_1K_1}{2\tilde v_1}}{v_1K_1}
\operatorname{SGN}(x-y,\tilde v_1 t),
\\[4pt]
[\hat{\phi}_{\mathrm{osc},1}(x,0),\hat{\theta}_{\mathrm{osc},2}(y,t)]
&=
\frac{i}{\pi}
\frac{g_1}{4\pi v_2K_2}
\frac{\pi v_1K_1}{2\tilde v_1}
\operatorname{SGN}(\tilde v_1 t,x-y),
\\[4pt]
[\hat{\phi}_{\mathrm{osc},2}(x,0),\hat{\theta}_{\mathrm{osc},1}(y,t)]
&=
\frac{i}{\pi}
\frac{g_2}{4\pi v_1K_1}
\frac{\pi v_2K_2}{2\tilde v_2}
\operatorname{SGN}(\tilde v_2 t,x-y),
\\[4pt]
[\hat{\phi}_{\mathrm{osc},2}(x,0),\hat{\theta}_{\mathrm{osc},2}(y,t)]
&=
\frac{i}{\pi}
\frac{\tilde v_2\frac{\pi v_2K_2}{2\tilde v_2}}{v_2K_2}
\operatorname{SGN}(x-y,\tilde v_2 t),
\\[6pt]
[\hat{\theta}_{\mathrm{osc},1}(x,0),\hat{\theta}_{\mathrm{osc},1}(y,t)]
&=
\frac{i}{\pi}
\left(\frac{1}{v_1K_1}\right)^2
\left[
\tilde v_1^2\frac{\pi v_1K_1}{2\tilde v_1}
\operatorname{SGN}(\tilde v_1 t,x-y)
+
\left(\frac{g_2}{4\pi}\right)^2
\frac{\pi v_2K_2}{2\tilde v_2}
\operatorname{SGN}(\tilde v_2 t,x-y)
\right],
\\[6pt]
[\hat{\theta}_{\mathrm{osc},1}(x,0),\hat{\theta}_{\mathrm{osc},2}(y,t)]
&=
\frac{i}{\pi}
\frac{1}{v_1K_1v_2K_2}
\left[
\frac{g_1\tilde v_1}{4\pi}
\frac{\pi v_1K_1}{2\tilde v_1}
\operatorname{SGN}(x-y,\tilde v_1 t)
+
\frac{g_2\tilde v_2}{4\pi}
\frac{\pi v_2K_2}{2\tilde v_2}
\operatorname{SGN}(x-y,\tilde v_2 t)
\right],
\\[6pt]
[\hat{\theta}_{\mathrm{osc},2}(x,0),\hat{\theta}_{\mathrm{osc},2}(y,t)]
&=
\frac{i}{\pi}
\left(\frac{1}{v_2K_2}\right)^2
\left[
\left(\frac{g_1}{4\pi}\right)^2
\frac{\pi v_1K_1}{2\tilde v_1}
\operatorname{SGN}(\tilde v_1 t,x-y)
+
\tilde v_2^2\frac{\pi v_2K_2}{2\tilde v_2}
\operatorname{SGN}(\tilde v_2 t,x-y)
\right].
\end{align}
\endgroup

\subsection{Density-Density Correlation Function for Open Boundary Conditions}
\label{subsec: densitydensityopen}

\begin{table}[!htbp]
\centering
\renewcommand{\arraystretch}{2.5}
\begin{tabular}{|c|c|}
\hline
$j$ &
$\left(\Gamma_{j,x}+\tilde{\sigma}\Gamma_{j,y}(t)\right)_{0}$
\\
\hline
$1$ &
$\displaystyle
2(1+\tilde{\sigma})\phi_{0,1}
-
\frac{2(\gamma_1+\pi\langle\hat{N}_1\rangle)}{L}
\left(x+\tilde{\sigma}y\right)
$
\\
\hline
$2$ &
$\displaystyle
2(1+\tilde{\sigma})\phi_{0,2}
-
\frac{2(\gamma_2+\pi\langle\hat{N}_2\rangle)}{L}
\left(x+\tilde{\sigma}y\right)
$
\\
\hline
\end{tabular}
\caption{Zero mode phase factors of the lowest order correlator for open boundary conditions in \refEq{lowestorderobccorrelator}. Only $\tilde{\sigma}=-1$ case shows translational invariance, in conjunction with the bulk limits in \refEq{bulklimitlowestorder}, whereas $\tilde{\sigma}=1$ cannot be interpreted as momentum shifts. The expectation values are taken with respect to the zero mode Hamiltonian in \refEq{zeromodeopen}, where $\phi_{0,j}=\pi/2$ for the realization in Ref. \cite{Batista2014}.}
\label{tableobczeromodes1}
\end{table}

\begin{table}[!htbp]
\centering
{\small
\renewcommand{\arraystretch}{1.35}
\[
\begin{array}{c|l}
j &
\left(\Gamma_{j,x}-\Gamma_{j,y}(t)\right)_{0}
\\
\hline
3 &
\begin{aligned}[t]
&
2Q(x-y)
+
\left(
\frac{g_{1}}{4\pi v_2K_2}
-1
\right)
\frac{(\gamma_1+\pi\langle\hat{N}_1\rangle)(x-y)}{L}
\\
&
-
\left(
1+
\frac{g_{2}}{4\pi v_1K_1}
\right)
\frac{(\gamma_2+\pi \langle\hat{N}_2\rangle)(x-y)}{L}
\\
&
-
\frac{\tilde{v}_1(\gamma_1+\pi\langle\hat{N}_1\rangle)t}{K_1L}
+
\frac{\tilde{v}_2(\gamma_2+\pi\langle\hat{N}_2\rangle)t}{K_2L}
\end{aligned}
\\[1.2em]
4 &
\begin{aligned}[t]
&
-2Q(x-y)
-
\left(
1+
\frac{g_{1}}{4\pi v_2K_2}
\right)
\frac{(\gamma_1+\pi\langle\hat{N}_1\rangle)(x-y)}{L}
\\
&
+
\left(
\frac{g_{2}}{4\pi v_1K_1}
-1
\right)
\frac{(\gamma_2+\pi\langle\hat{N}_2\rangle)(x-y)}{L}
\\
&
+
\frac{\tilde{v}_1(\gamma_1+\pi\langle\hat{N}_1\rangle)t}{K_1L}
-
\frac{\tilde{v}_2(\gamma_2+\pi\langle\hat{N}_2\rangle)t}{K_2L}
\end{aligned}
\\[1.2em]
5 &
\begin{aligned}[t]
&
2Q(x-y)
+
\left(
\frac{g_{1}}{4\pi v_2K_2}
-1
\right)
\frac{(\gamma_1+\pi\langle\hat{N}_1\rangle)(x-y)}{L}
\\
&
+
\left(
1-
\frac{g_{2}}{4\pi v_1K_1}
\right)
\frac{(\gamma_2+\pi\langle\hat{N}_2\rangle)(x-y)}{L}
\\
&
-
\frac{\tilde{v}_1(\gamma_1+\pi\langle\hat{N}_1\rangle)t}{K_1L}
+
\frac{\tilde{v}_2(\gamma_2+\pi\langle\hat{N}_2\rangle)t}{K_2L}
\end{aligned}
\\[1.2em]
6 &
\begin{aligned}[t]
&
-2Q(x-y)
-
\left(
1+
\frac{g_{1}}{4\pi v_2K_2}
\right)
\frac{(\gamma_1+\pi \langle\hat{N}_1\rangle)(x-y)}{L}
\\
&
+
\left(
1+
\frac{g_{2}}{4\pi v_1K_1}
\right)
\frac{(\gamma_2+\pi\langle\hat{N}_2\rangle)(x-y)}{L}
\\
&
+
\frac{\tilde{v}_1(\gamma_1+\pi\langle\hat{N}_1\rangle)t}{K_1L}
-
\frac{\tilde{v}_2(\gamma_2+\pi\langle\hat{N}_2\rangle)t}{K_2L}
\end{aligned}
\end{array}
\]
}
\caption{Zero mode phase factors of the subleading contributions for the open boundary correlation function in \refEq{higherorderobc}. In contrast to \refTab{tableobczeromodes1} the phase factors can be interpreted as energy and momentum shifts, depending also on the zero mode expectation value according to \refEq{zeromodeopen}.}
\label{tableobczeromodes2}
\end{table}

In the following section, we define the phase factors of the density-density correlation function in \refEq{correlatorobc}, whereas the decay coefficients can be found already in the maintext, in contrast to the periodic case.
The calculations are analogous to \refSec{subsec: densitydensityperiodicpbc}, where \refEq{Dirichletphi} and \refEq{Dirichletpi} have to be inserted now into \refEq{definitionsgamma0}. The zero mode parts are shown in \refTab{tableobczeromodes1} and \refTab{tableobczeromodes2} for the leading and subleading contributions of \refEq{correlatorobc}, respectively. Due to the loss of translational invariance, the lattice positions $x$ and $y$ acquire different phase factors that are generally not the additive inverse of each other, whereas expectation values of zero modes are taken with respect to \refEq{zeromodeopen}.

\subsubsection{Commutator Phases}\label{subsec: Commutator Phasesobc}
For the calculation of the commutator phases $\left[\Gamma_{j,x},\Gamma_{j,y}(t)\right]$ in \refEq{correlatorobc} we split the mode decompositions in \refEq{Dirichletphi} and \refEq{Dirichletpi} again additively into zero modes and finite mode components. For the latter, the following abbreviations for a linear combination of signum functions are useful,
\begin{align}
\mathcal A_{1,2}(x,y;t)
&\equiv
\operatorname{sgn}(\tilde{v}_{1,2}t+x-y)
+\operatorname{sgn}(\tilde{v}_{1,2}t-x+y)
-\operatorname{sgn}(\tilde{v}_{1,2}t+x+y)
-\operatorname{sgn}(\tilde{v}_{1,2}t-x-y),\nonumber
\\
\mathcal B_{1,2}(x,y;t)
&\equiv
\operatorname{sgn}(\tilde{v}_{1,2}t+x+y)
+\operatorname{sgn}(\tilde{v}_{1,2}t+x-y)
+\operatorname{sgn}(\tilde{v}_{1,2}t-x+y)
+\operatorname{sgn}(\tilde{v}_{1,2}t-x-y)
-\frac{4\tilde{v}_{1,2}t}{L},
\\
\mathcal C_{1,2}(x,y;t)
&\equiv
\operatorname{sgn}(x-y+\tilde{v}_{1,2}t)
+\operatorname{sgn}(x-y-\tilde{v}_{1,2}t)
-\frac{2(x-y)}{L}.\nonumber
\end{align}
Consequently, we denote species diagonal commutators by $F_{i=j}^{\rm OBC}$, 
\begin{align}
F_{11}^{\rm OBC}
&\equiv
[\hat{\phi}_{\rm osc,1}(x,0),\hat{\phi}_{\rm osc,1}(y,t)]
=
\frac{i\pi K_1v_1}{4\tilde{v}_{1}}\,
\mathcal A_{1}(x,y;t),
\\
F_{33}^{\rm OBC}
&\equiv
[\hat{\phi}_{\rm osc,2}(x,0),\hat{\phi}_{\rm osc,2}(y,t)]
=
\frac{i\pi K_2v_2}{4\tilde{v}_{2}}\,
\mathcal A_{2}(x,y;t),
\\
F_{22}^{\rm OBC}
&\equiv
[\hat{\theta}_{\rm osc,1}(x,0),\hat{\theta}_{\rm osc,1}(y,t)]
=
i\,\frac{g_2^2K_2v_2}{64\pi K_1^2v_1^2\tilde{v}_{2}}\,
\mathcal A_{2}(x,y;t)
+
i\,\frac{\pi \tilde{v}_{1}}{4K_1v_1}\,
\mathcal B_{1}(x,y;t),\nonumber
\\
F_{44}^{\rm OBC}
&\equiv
[\hat{\theta}_{\rm osc,2}(x,0),\hat{\theta}_{\rm osc,2}(y,t)]
=
i\,\frac{g_1^2K_1v_1}{64\pi K_2^2v_2^2\tilde{v}_{1}}\,
\mathcal A_{1}(x,y;t)
+
i\,\frac{\pi \tilde{v}_{2}}{4K_2v_2}\,
\mathcal B_{2}(x,y;t),\nonumber
\end{align}
and commutators between the species by  
$G_{i\neq j}^{\rm OBC}$,  
\begin{align}
G_{12}^{\rm OBC}
&\equiv
[\hat{\phi}_{\rm osc,1}(x,0),\hat{\theta}_{\rm osc,1}(y,t)]
+
[\hat{\theta}_{\rm osc,1}(x,0),\hat{\phi}_{\rm osc,1}(y,t)]
=
-\frac{i\pi}{2}\,
\mathcal C_{1}(x,y;t),
\\
G_{34}^{\rm OBC}
&\equiv
[\hat{\phi}_{\rm osc,2}(x,0),\hat{\theta}_{\rm osc,2}(y,t)]
+
[\hat{\theta}_{\rm osc,2}(x,0),\hat{\phi}_{\rm osc,2}(y,t)]
=
-\frac{i\pi}{2}\,
\mathcal C_{2}(x,y;t),
\\
G_{14}^{\rm OBC}
&\equiv
[\hat{\phi}_{\rm osc,1}(x,0),\hat{\theta}_{\rm osc,2}(y,t)]
+
[\hat{\theta}_{\rm osc,2}(x,0),\hat{\phi}_{\rm osc,1}(y,t)]
=
-\,i\,\frac{g_1K_1v_1}{8K_2v_2\tilde{v}_{1}}\,
\mathcal A_{1}(x,y;t),
\\
G_{23}^{\rm OBC}
&\equiv
[\hat{\theta}_{\rm osc,1}(x,0),\hat{\phi}_{\rm osc,2}(y,t)]
+
[\hat{\phi}_{\rm osc,2}(x,0),\hat{\theta}_{\rm osc,1}(y,t)]
=
-\,i\,\frac{g_2K_2v_2}{8K_1v_1\tilde{v}_{2}}\,
\mathcal A_{2}(x,y;t),
\\
G_{13}^{\rm OBC}
&\equiv
[\hat{\phi}_{\rm osc,1},\hat{\phi}_{\rm osc,2}]
+
[\hat{\phi}_{\rm osc,2},\hat{\phi}_{\rm osc,1}]
=0,
\\
G_{24}^{\rm OBC}
&\equiv
[\hat{\theta}_{\rm osc,1}(x,0),\hat{\theta}_{\rm osc,2}(y,t)]
+
[\hat{\theta}_{\rm osc,2}(x,0),\hat{\theta}_{\rm osc,1}(y,t)]
=
i\,\frac{g_1}{8K_2v_2}\,
\mathcal C_{1}(x,y;t)
+
i\,\frac{g_2}{8K_1v_1}\,
\mathcal C_{2}(x,y;t).
\end{align}

The oscillator part of the commutator phases in \refEq{correlatorobc} is then finally obtained from the following expressions,
\begin{align}
[\Gamma_1(x,0),\Gamma_1(y,t)]_{\rm osc}^{\rm OBC}
&=
4F_{11}^{\rm OBC},
\\[4pt]
[\Gamma_2(x,0),\Gamma_2(y,t)]_{\rm osc}^{\rm OBC}
&=
4F_{33}^{\rm OBC},
\\[6pt]
[\Gamma_3(x,0),\Gamma_3(y,t)]_{\rm osc}^{\rm OBC}
&=
F_{11}^{\rm OBC}
+F_{22}^{\rm OBC}
+F_{33}^{\rm OBC}
+F_{44}^{\rm OBC}
-G_{12}^{\rm OBC}
+G_{14}^{\rm OBC}
-G_{23}^{\rm OBC}
-G_{24}^{\rm OBC}
+G_{34}^{\rm OBC},
\\[6pt]
[\Gamma_4(x,0),\Gamma_4(y,t)]_{\rm osc}^{\rm OBC}
&=
F_{11}^{\rm OBC}
+F_{22}^{\rm OBC}
+F_{33}^{\rm OBC}
+F_{44}^{\rm OBC}
+G_{12}^{\rm OBC}
-G_{14}^{\rm OBC}
+G_{23}^{\rm OBC}
-G_{24}^{\rm OBC}
-G_{34}^{\rm OBC},
\\[6pt]
[\Gamma_5(x,0),\Gamma_5(y,t)]_{\rm osc}^{\rm OBC}
&=
F_{11}^{\rm OBC}
+F_{22}^{\rm OBC}
+F_{33}^{\rm OBC}
+F_{44}^{\rm OBC}
-G_{12}^{\rm OBC}
+G_{14}^{\rm OBC}
+G_{23}^{\rm OBC}
-G_{24}^{\rm OBC}
-G_{34}^{\rm OBC},
\\[6pt]
[\Gamma_6(x,0),\Gamma_6(y,t)]_{\rm osc}^{\rm OBC}
&=
F_{11}^{\rm OBC}
+F_{22}^{\rm OBC}
+F_{33}^{\rm OBC}
+F_{44}^{\rm OBC}
+G_{12}^{\rm OBC}
-G_{14}^{\rm OBC}
-G_{23}^{\rm OBC}
-G_{24}^{\rm OBC}
+G_{34}^{\rm OBC}.
\end{align}
with analogous interpretation to \refSec{subsec: Commutator Phasesperiodic}.
\\
\\
Finally, we discuss the zero mode part of the commutator phases in \refEq{correlatorobc}. To this end we define the function $\mathcal{Z}^{\rm OBC}_0(x,y;t)$ that appears ubiquitously,
\begin{align}
\mathcal{Z}^{\rm OBC}_0(x,y;t)
&\equiv
\frac{i\pi t}{L}
\left(
\frac{\tilde{v}_{1}}{K_1}
+
\frac{\tilde{v}_{2}}{K_2}
\right)
-
\frac{i(x-y)}{4L}
\left(
\frac{g_2}{v_1K_1}
+
\frac{g_1}{v_2K_2}
\right),
\label{universalzeromodefactorincommutatorsobc}
\end{align}
Consequently, the different zero mode contributions $[\Gamma_j(x,0),\Gamma_j(y,t)]_{\rm zero}^{\rm OBC}$  per index $j$ are listed in \refTab{tab:Gamma_zero_OBC}

\begin{table}[!htbp]
\centering
\renewcommand{\arraystretch}{1.3}
\setlength{\tabcolsep}{8pt}
\begin{tabular}{c|l}
$j$ & $[\Gamma_j(x,0),\Gamma_j(y,t)]_{0}^{\rm OBC}$ \\
\hline

1 &
$0$
\\[4pt]

2 &
$0$
\\[4pt]

3 &
$\mathcal Z^{\rm OBC}_0(x,y;t)$
\\[4pt]

4 &
$\mathcal Z^{\rm OBC}_0(x,y;t)$
\\[4pt]

5 &
$\displaystyle \mathcal Z^{\rm OBC}_0(x,y;t)+\frac{2i\pi}{L}(x-y)$
\\[6pt]

6 &
$\displaystyle \mathcal Z^{\rm OBC}_0(x,y;t)-\frac{2i\pi}{L}(x-y)$
\\
\hline
\end{tabular}
\caption{Zero-mode contribution of commutator phases for the open-boundary density-density correlator in \refEq{correlatorobc}. Here,
$\mathcal Z^{\rm OBC}_0(x,y;t)$ is defined in \refEq{universalzeromodefactorincommutatorsobc}.}
\label{tab:Gamma_zero_OBC}
\end{table}

\clearpage

\clearpage
\bibliographystyle{apsrev4-2}

%

\end{document}